\documentclass{emulateapj}

\usepackage{natbib,graphicx,textcomp,lscape,amsmath,amsthm,ulem, subfigure,color}

\newcommand{\ecosw}{\ensuremath{\sqrt{e}\cos{\omega_{\star}}}}
\newcommand{\esinw}{\ensuremath{\sqrt{e}\sin{\omega_{\star}}}}
\newcommand{\bjdtdb}{\ensuremath{\rm {BJD_{TDB}}}}

\newcommand{\feh}{\ensuremath{[\mbox{Fe}/\mbox{H}]}}

\newcommand{\caii}{\ion{Ca}{2} H \& K}
\newcommand{\logr}{log(\ensuremath{R'_{\mbox{\scriptsize HK}}})}
\newcommand{\vsini}{$v \sin i$}
\newcommand{\sval}{\ensuremath{S_{\mbox{\scriptsize HK}}}}

\newcommand{\mj}{\ensuremath{\,{\rm M}_{\rm J}}}

\newcommand{\ms}{m s$^{-1}$}

\begin{document}

\title{Friends of Hot Jupiters I:  A Radial Velocity Search for Massive, Long-Period Companions to Close-In Gas Giant Planets} 

\author{
Heather A. Knutson\altaffilmark{1,2}, Benjamin J. Fulton\altaffilmark{3}, Benjamin T. Montet\altaffilmark{4,5}, Melodie Kao\altaffilmark{4}, Henry Ngo\altaffilmark{1}, Andrew W. Howard\altaffilmark{3}, Justin R. Crepp\altaffilmark{6}, Sasha Hinkley\altaffilmark{4,7}, Gaspar \' A. Bakos\altaffilmark{8,9}, Konstantin Batygin\altaffilmark{5}, John Asher Johnson\altaffilmark{1,5}, Timothy D. Morton\altaffilmark{4,9}, Philip S. Muirhead\altaffilmark{4,10}
} 

\altaffiltext{1}{Division of Geological and Planetary Sciences, California Institute of Technology, Pasadena, CA 91125, USA} 
\altaffiltext{2}{hknutson@caltech.edu}
\altaffiltext{3}{Institute for Astronomy, University of Hawaii at Manoa, Honolulu, HI, USA}
\altaffiltext{4}{Cahill Center for Astronomy and Astrophysics, California Institute of Technology, 1200 E. California Blvd., MC 249-17, Pasadena, CA 91125, USA}
\altaffiltext{5}{Harvard-Smithsonian Center for Astrophysics, Cambridge MA, USA}
\altaffiltext{6}{University of Notre Dame, Department of Physics, Notre Dame, IN, USA}
\altaffiltext{7}{NSF Astronomy and Astrophysics Postdoctoral Fellow}
\altaffiltext{8}{Alfred P. Sloan Fellow, Packard Fellow}
\altaffiltext{9}{Princeton University, Department of Astrophysical Sciences, Princeton, NJ, USA}
\altaffiltext{10}{Hubble Fellow; Boston University, Department of Astronomy, Boston, MA, USA}

\begin{abstract}

In this paper we search for distant massive companions to known transiting gas giant planets that may have influenced the dynamical evolution of these systems.  We present new radial velocity observations for a sample of 51 planets obtained using the Keck HIRES instrument, and find statistically significant accelerations in fifteen systems.  Six of these systems have no previously reported accelerations in the published literature: HAT-P-10, HAT-P-22, HAT-P-29, HAT-P-32, WASP-10, and XO-2.  We combine our radial velocity fits with Keck NIRC2 adaptive optics (AO) imaging data to place constraints on the allowed masses and orbital periods of the companions responsible for the detected accelerations.  The estimated masses of the companions range between $1-500$~M$_{Jup}$, with orbital semi-major axes typically between $1-75$ AU.  A significant majority of the companions detected by our survey are constrained to have minimum masses comparable to or larger than those of the transiting planets in these systems, making them candidates for influencing the orbital evolution of the inner gas giant.  We estimate a total occurrence rate of $51\pm10\%$ for companions with masses between $1-13$~M$_{Jup}$ and orbital semi-major axes between $1-20$ AU in our sample.  We find no statistically significant difference between the frequency of companions to transiting planets with misaligned or eccentric orbits and those with well-aligned, circular orbits.  We combine our expanded sample of radial velocity measurements with constraints from transit and secondary eclipse observations to provide improved measurements of the physical and orbital characteristics of all of the planets included in our survey.  
\end{abstract}

\keywords{binaries: eclipsing --- planetary systems --- techniques: radial velocity, adaptive optics}

\section{Introduction}\label{intro}

Observations of exoplanetary systems offer a unique window into the processes that drive planet formation and migration.  The short-period, gas giant planets known as hot Jupiters pose a particular challenge for planet formation models, as we know that they could not have formed at their present-day locations but instead must have migrated inward from beyond the ice line \citep[e.g.,][]{lin96}.  Hot Jupiter migration models can be broadly divided into several classes, including disk-driven migration, binary star-planet interactions, and planet-planet interactions.  In the simplest disk migration models, including both Type I and II migration, we expect the resulting short-period planets to have largely circular orbits that are well-aligned relative to the stars spin axis \citep[e.g.,][]{goldreich80,lin86,ward97,tanaka02}.  In contrast to this result, migration mechanisms involving multi-body interactions such as Kozai migration \citep[e.g.,][]{wu03,malmberg07,fabrycky07,naoz12,teyssandier13}, which requires a distant stellar companion, planet-planet scattering \citep[e.g.,][]{chatterjee08,nagasawa08}, and secular chaotic excursions \citep{wu11,lithwick13} frequently produce close-in planets with misaligned and/or eccentric orbits.  

There are currently two systems (HD 80606b and 16 Cyg Bb) where there is clear evidence for orbital evolution of an eccentric Jovian-mass exoplanet due to interactions with a distant stellar companion \citep{holman97,wu03}.  There are also several known two-planet systems where the inner gas giant exchanges eccentricity and angular momentum with a massive outer planetary companion \citep[e.g.,][]{kane14a,kane14b}.  The recent discovery of an eccentric, short-period Jupiter in the young Hyades cluster also appears to be consistent with high-eccentricity migration mechanisms \citep{quinn13}, although uneven irradiation of disk gaps might excite the eccentricities of Jovian mass planets \citep{goldreich03,tsang13}.  \citet{juric08} proposed that planet-planet scattering could explain the high average eccentricities of the gas giant planets detected using the radial velocity technique.  However, \citet{dawson12} argued more recently that the lack of high-eccentricity Jupiters among the \textit{Kepler} transiting planet candidates places a limit on the relative number of planets that migrate via high-eccentricity mechanisms.  

Measurements of the spin-orbit alignments of transiting hot Jupiters via the Rossiter-McLaughlin effect \citep{winn05} indicate that almost half of the hot Jupiters surveyed to date have orbits that are significantly misaligned with respect to the starÕs spin axis \citep[e.g.,][]{winn10b,hebrard11,albrecht12b}.  Based on the arguments given above, this would seem to favor migration models involving either a second star or multiple planets \citep{morton11,li13}.  However, in the disk-driven migration case a distant stellar companion could also tilt the primordial disk, resulting in an alternative channel for spin-orbit misalignment \citep{batygin12,batygin13}.  The recent discovery of a short-period misaligned hot Jupiter orbiting a T Tauri star \citep{vaneyken12,barnes13} and a coplanar misaligned multi-planet system \citep{huber13} both provide strong evidence that such primordial disk misalignments do indeed occur in practice, although \citet{kaib11} argue that multi-planet systems could also be tilted by a stellar companion after the dissipation of the disk.  If we relax the assumption that the planet must be coplanar with the disk, interactions between the planet and the disk could also result in an eccentric, misaligned orbit \citep{teyssandier13a}. 

If multi-body dynamics play an important role in the orbital evolution of hot Jupiters, then such systems must necessarily include massive planetary or stellar companions that drive this dynamical evolution.  The most recent statistics from the \textit{Kepler} mission and radial velocity surveys indicate that many low-mass candidate planets exist in multi-planet systems \citep{tremaine12,batalha13,steffen13}, but the candidate hot Jupiters detected by the \textit{Kepler} survey rarely have nearby companions \citep{steffen12}.  This distinction also appears in measurements of spin-orbit alignment for the two types of systems, as the majority of multi-planet systems with published Rossiter measurements appear to be well-aligned with their host stars \citep{sanchis12,hirano12,albrecht13} while hot Jupiters are frequently misaligned \citep[e.g.,][]{albrecht12b}.  These two lines of evidence suggest that hot Jupiters likely formed via different evolutionary channels than the compact, low-mass multi-planet systems detected by \textit{Kepler}, but the underlying cause of this divergence is poorly understood.

\begin{figure}[ht]
\epsscale{1.2}
\plotone{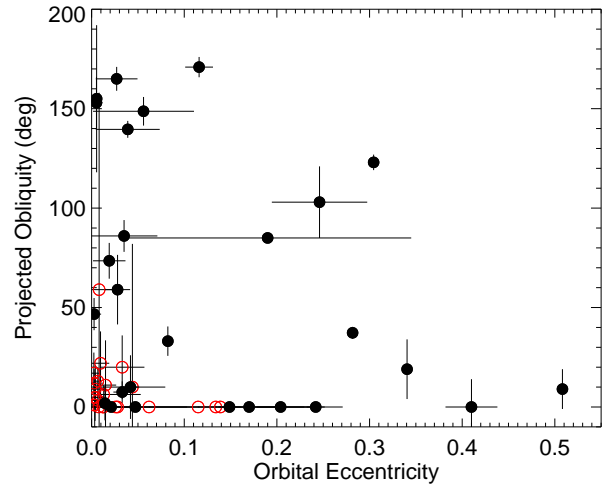}
\centering
\caption{A comparison of the projected orbital obliquities and eccentricities of the transiting gas giant planets in our two samples.  The sample of misaligned and/or eccentric planets is shown as black filled circles, while the control sample of planets with apparently circular and well-aligned orbits is shown as open red circles.  We plot the fourteen planets without measured obliquities along the x axis.}
\label{ecc_vs_spin}
\end{figure}

\begin{figure}[ht]
\epsscale{1.2}
\plotone{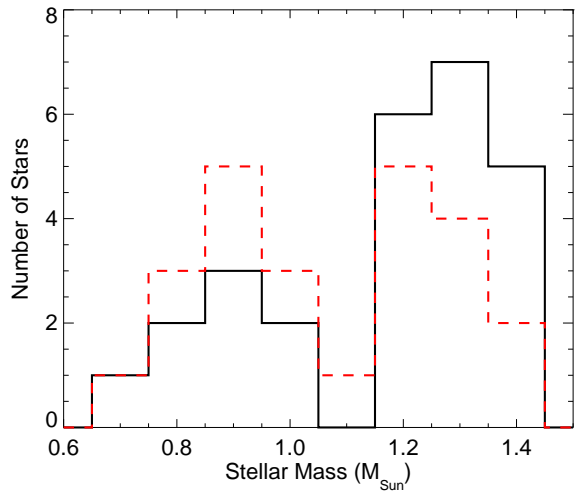}
\centering
\caption{Histogram of the stellar masses in our two samples.  The sample of misaligned and/or eccentric planets is shown as a black solid line, while the control sample of  planets with apparently circular and well-aligned orbits is shown as the red dashed line.  Masses for individual stars and associated references are listed in Table \ref{stellar_parameters}.}
\label{mass_hist}
\end{figure}

\begin{deluxetable*}{lllllll}
\tablecaption{Stellar Parameters \label{stellar_parameters}}
\tablewidth{0pt}
\tablehead{
\colhead{Star} & \colhead{Mass (M$_{\sun}$)} & \colhead{\feh} & \colhead{Sample\tablenotemark{a}} & \colhead{Reference} \\
}
\startdata
GJ436 & $0.452\pm0.013$ & $-0.03\pm0.20$ & Misaligned &  \citet{vonbraun12,bonfils05} \\ 
HAT-P-2 & $1.36\pm0.04$ & $0.14\pm0.08$ & Misaligned &  \citet{pal10} \\ 
HAT-P-4 & $1.26\pm0.1$ & $0.20\pm0.08$ & Control & \citet{winn11,torres12} \\ 
HAT-P-6 & $1.29\pm0.06$ & $-0.11\pm0.08$ & Misaligned & \citet{noyes08,torres12} \\ 
HAT-P-7 & $1.361\pm0.021$ & $0.15\pm0.08$ & Misaligned & \citet{vaneylen12,torres12} \\ 
HAT-P-8 & $1.192\pm0.075$ & $-0.04\pm0.08$ & Control & \citet{mancini13,torres12} \\ 
HAT-P-10 & $0.83\pm0.03$ & $0.25\pm0.07$ & Control & \citet{bakos09a,torres12} \\ 
HAT-P-11 & $0.81\pm0.03$ & $0.33\pm0.07$ & Misaligned & \citet{bakos10,torres12} \\ 
HAT-P-12 & $0.733\pm0.018$ & $-0.29\pm0.05$ & Control & \citet{hartman09} \\ 
HAT-P-13 & $1.320\pm0.062$ & $0.46\pm0.07$ & Misaligned & \citet{southworth12,torres12} \\ 
HAT-P-14 & $1.418\pm0.054$ & $0.07\pm0.08$ & Misaligned & \citet{southworth12b,torres12} \\ 
HAT-P-15 & $1.013\pm0.043$ & $0.31\pm0.08$ & Misaligned & \citet{kovacs10,torres12} \\ 
HAT-P-16 & $1.218\pm0.039$ & $0.12\pm0.08$ & Misaligned & \citet{buchhave10,torres12} \\ 
HAT-P-17 & $0.857\pm0.039$ & $0.06\pm0.08$ & Misaligned & \citet{howard12,torres12} \\ 
HAT-P-18 & $0.77\pm0.03$ & $0.14\pm0.08$ & Control & \citet{hartman11a,torres12} \\ 
HAT-P-20 & $0.756\pm0.028$ & $0.26\pm0.11$ & Misaligned & \citet{bakos11,torres12} \\ 
HAT-P-22 & $0.916\pm0.035$ & $0.29\pm0.08$ & Control & \citet{bakos11,torres12} \\ 
HAT-P-24 & $1.191\pm0.042$ & $-0.21\pm0.08$ & Control & \citet{kipping10,torres12} \\ 
HAT-P-26 & $0.816\pm0.033$ & $0.10\pm0.08$ & Control & \citet{hartman11b,torres12} \\ 
HAT-P-29 & $1.207\pm0.046$ & $0.14\pm0.08$ & Control & \citet{buchhave11,torres12} \\ 
HAT-P-30 & $1.242\pm0.041$ & $0.13\pm0.08$ & Misaligned & \citet{johnson11} \\ 
HAT-P-31 & $1.218\pm0.07$ & $0.15\pm0.08$ & Misaligned & \citet{kipping11} \\ 
HAT-P-32 & $1.16\pm0.04$ & $-0.04\pm0.08$ & Misaligned & \citet{hartman11c} \\ 
HAT-P-33 & $1.38\pm0.04$ & $0.07\pm0.08$ & Control & \citet{hartman11c} \\ 
HAT-P-34 & $1.392\pm0.047$ & $0.21\pm0.10$ & Misaligned & \citet{bakos12} \\ 
HD149026 & $1.345\pm0.020$ & $0.24\pm0.07$ & Control & \citet{carter09,torres12} \\ 
TrES-2 & $0.94\pm0.05$ & $-0.01\pm0.08$ & Control & \citet{barclay12,torres12} \\ 
TrES-3 & $0.928\pm0.038$ & $-0.20\pm0.07$ & Misaligned & \citet{sozzetti09,torres12} \\ 
TrES-4 & $1.339\pm0.086$ & $0.14\pm0.09$ & Control & \citet{sozzetti09} \\ 
WASP-1 & $1.27\pm0.05$ & $0.14\pm0.07$ & Control & \citet{southworth12b,torres12} \\ 
WASP-2 & $0.85\pm0.05$ & $0.06\pm0.07$ & Misaligned & \citet{southworth12b,torres12} \\ 
WASP-3 & $1.20\pm0.01$ & $-0.06\pm0.08$ & Control & \citet{pollacco08,torres12} \\ 
WASP-4 & $0.92\pm0.07$ & $0.0\pm0.2$ & Control & \citet{doyle13,wilson08} \\ 
WASP-7 & $1.34\pm0.09$ & $0.0\pm0.1$ & Misaligned & \citet{doyle13,hellier08} \\ 
WASP-8 & $1.04\pm0.08$ & $0.17\pm0.07$ & Misaligned & \citet{doyle13,queloz10} \\ 
WASP-10 & $0.75\pm0.03$ & $0.05\pm0.08$ & Misaligned & \citet{johnson09b,torres12} \\ 
WASP-12 & $1.38\pm0.19$ & $0.07\pm0.07$ & Misaligned &  \citet{southworth12b,torres12} \\ 
WASP-14 & $1.35\pm0.12$ & $-0.13\pm0.08$ & Misaligned & \citet{southworth12b,torres12} \\ 
WASP-15 & $1.305\pm0.051$ & $0.0\pm0.1$ & Misaligned & \citet{southworth13} \\ 
WASP-16 & $0.98\pm0.05$ & $0.07\pm0.10$ & Control & \citet{southworth13} \\ 
WASP-17 & $1.286\pm0.079$ & $-0.02\pm0.09$ & Misaligned & \citet{southworth12,torres12} \\ 
WASP-18 & $1.28\pm0.09$ & $0.11\pm0.08$ & Control & \citet{doyle13,torres12} \\ 
WASP-19 & $0.935\pm0.041$ & $0.15\pm0.07$ & Control &  \citet{mancini13b,torres12} \\ 
WASP-22 & $1.109\pm0.026$ & $0.05\pm0.08$ & Control & \citet{anderson11} \\ 
WASP-24 & $1.184\pm0.027$ & $-0.02\pm0.10$ & Control & \citet{street10,torres12} \\ 
WASP-34 & $1.01\pm0.07$ & $-0.02\pm0.10$ & Control & \citet{smalley11} \\ 
WASP-38 & $1.23\pm0.04$ & $-0.02\pm0.10$ & Misaligned & \citet{brown12b,torres12} \\ 
XO-2 & $0.98\pm0.02$ & $0.35\pm0.08$ & Control & \citet{burke07,torres12} \\ 
XO-3 & $1.213\pm0.066$ & $-0.05\pm0.08$ & Misaligned & \citet{winn08b,torres12} \\ 
XO-4 & $1.32\pm0.02$ & $-0.03\pm0.08$ & Misaligned & \citet{mccullough08,torres12} \\ 
XO-5 & $0.88\pm0.03$ & $0.05\pm0.06$ & Control & \citet{pal09} \\
\enddata
\tablenotetext{a}{The misaligned sample consists of planets with either eccentric or misaligned orbits, while the control sample contains planets that appear to have circular and/or well-aligned orbits.}
\end{deluxetable*}

Although massive, long-period companions may play a significant role in shaping the observed properties of hot Jupiters,  most confirmed transiting planet systems have only received a handful of follow-up radial velocity measurements immediately after the initial discovery of the transit signal \citep{madhu09,pont11,husnoo12}.  Observations of field stars indicate that more than half of solar type stars exist in binary or multiple systems \citep{duquennoy91,raghavan10}; if exoplanetary systems follow the same pattern, then it is possible that many of these systems have currently unknown low-mass stellar companions.  This paper is the first in a three-part series describing a search for distant stellar and massive planetary companions to a sample of 51 known  short period gas giant planets.  We focus here on long-term radial velocity monitoring, while in the second and third paper we will present complementary K-band adaptive optics (AO) imaging and high-resolution K-band spectroscopy of our target stars, respectively.  By combining multiple techniques, we ensure maximum sensitivity to companions spanning a broad range of orbital separations.  Radial velocity monitoring can detect gas giant planets at distances of up to 5-10 AU and stellar companions out to larger distances, while infrared spectroscopy is sensitive to low-mass stellar companions within 0.5\arcsec~(approximately 50 AU for most of the stars in our sample), and K-band AO imaging can detect stellar companions at distances between 50-200 AU  

\begin{deluxetable*}{l c c c c c c c}
\tablecaption{Summary of Radial Velocity Observations \label{rv_obs}}
\renewcommand{\arraystretch}{0.7}
\tablewidth{0pt}
\tablehead{
\colhead{Star} & \colhead{$N_{\rm CPS}$\tablenotemark{a}} & \colhead{Start date} & \colhead{End date} & \colhead{Duration} & \colhead{$N_{\rm data}$\tablenotemark{b}} & \colhead{Sample\tablenotemark{c}} & \colhead{Ref.} \\
\colhead{} & \colhead{} & \colhead{UTC} & \colhead{UTC} & \colhead{days} & \colhead{} & \colhead{} & \colhead{}
}
\startdata
GJ436 & 113 & 2000-01-08 & 2012-12-04 & 4714 & 2 & Misaligned &  \\ 
HAT-P-2 & 40 & 2006-09-04 & 2013-08-02 & 2524 & 1 & Misaligned &  \\ 
HAT-P-4 & 23 & 2007-03-27 & 2012-07-04 & 1926 & 1 & Control &  \\ 
HAT-P-6 & 25 & 2006-10-14 & 2013-07-24 & 2475 & 1 & Misaligned &  \\ 
HAT-P-7 & 43 & 2007-08-24 & 2013-07-12 & 2149 & 1 & Misaligned &  \\ 
HAT-P-8 & 16 & 2007-08-24 & 2013-08-28 & 2196 & 1 & Control &  \\ 
HAT-P-10 & 13 & 2008-03-22 & 2012-09-25 & 1648 & 1 & Control &  \\ 
HAT-P-11 & 77 & 2007-08-23 & 2013-07-13 & 2151 & 1 & Misaligned &  \\ 
HAT-P-12 & 23 & 2007-03-27 & 2013-02-21 & 2158 & 1 & Control  &  \\ 
HAT-P-13 & 63 & 2008-03-23 & 2013-02-03 & 1778 & 1 & Misaligned &  \\ 
HAT-P-14 & 17 & 2008-05-16 & 2012-08-08 & 1545 & 1 & Misaligned &  \\ 
HAT-P-15 & 28 & 2007-08-24 & 2012-09-25 & 1859 & 1 & Misaligned &  \\ 
HAT-P-16 & 10 & 2009-07-04 & 2012-07-25 & 1117 & 1 & Misaligned &  \\ 
HAT-P-17 & 47 & 2007-10-23 & 2013-08-28 & 2136 & 1 & Misaligned &  \\ 
HAT-P-18 & 31 & 2007-10-24 & 2012-06-01 & 1682 & 1 & Control &  \\ 
HAT-P-20 & 13 & 2009-04-13 & 2012-12-04 & 1331 & 1 & Misaligned &  \\ 
HAT-P-22 & 18 & 2009-04-07 & 2012-12-28 & 1361 & 1 & Control &  \\ 
HAT-P-24 & 24 & 2009-04-07 & 2012-12-04 & 1337 & 1 & Control &  \\ 
HAT-P-26 & 26 & 2009-12-27 & 2013-02-04 & 1135 & 1 & Control & 1 \\ 
HAT-P-29 & 11 & 2010-09-26 & 2012-08-25 & 699 & 1 & Control &  \\ 
HAT-P-30 & 19 & 2010-04-27 & 2012-12-04 & 952 & 1 & Misaligned &  \\ 
HAT-P-31 & 11 & 2009-08-08 & 2012-07-05 & 1062 & 3 & Misaligned & 2\\ 
HAT-P-32 & 30 & 2007-08-24 & 2012-08-25 & 1828 & 1 & Misaligned & \\ 
HAT-P-33 & 26 & 2008-09-18 & 2012-12-04 & 1538 & 1 & Control & \\ 
HAT-P-34 & 17 & 2010-06-26 & 2012-08-14 & 780 & 1 & Misaligned & \\ 
HD149026 & 43 & 2005-02-27 & 2013-08-27 & 3103 & 1 & Control  &  \\ 
TrES-2 & 19 & 2007-04-26 & 2012-10-08 & 1992 & 1 & Control & \\ 
TrES-3 & 8 & 2007-03-27 & 2012-07-25 & 1947 & 2 & Misaligned &  3\\ 
TRES-4 & 6 & 2007-03-27 & 2012-08-01 & 1954 & 3 & Control & 4,5\\ 
WASP-1 & 10 & 2006-09-01 & 2012-08-24 & 2184 & 5 & Control &  6,7,8\\ 
WASP-2 & 6 & 2006-09-03 & 2012-09-09 & 2197 & 6 & Misaligned & 6,9,10 \\ 
WASP-3 & 15 & 2007-07-05 & 2012-08-08 & 1861 & 3 & Control & 11,12,13\\ 
WASP-4 & 5 & 2007-09-16 & 2013-08-27 & 2172 & 4 & Control & 9,10,14\\ 
WASP-7 & 18 & 2007-08-17 & 2012-10-08 & 1879 & 5 & Misaligned & 9,10,15,16\\ 
WASP-8 & 9 & 2007-11-29 & 2013-08-27 & 2099 & 3 & Misaligned & 17\\ 
WASP-10 & 9 & 2007-08-28 & 2013-08-28 & 2192 & 2 & Misaligned & 18\tablenotemark{d}\\ 
WASP-12 & 30 & 2008-02-12 & 2013-12-11 & 2129 & 3 & Misaligned & 19,20\\ 
WASP-14 & 9 & 2007-12-27 & 2012-03-05 & 1530 & 6 & Misaligned & 20,21,22\\ 
WASP-15 & 2 & 2008-03-06 & 2012-07-01 & 1578 & 2 & Misaligned & 23\\ 
WASP-16 & 4 & 2008-03-10 & 2012-07-01 & 1574 & 3 & Control & 24,25\\ 
WASP-17 & 5 & 2007-08-17 & 2012-09-09 & 1850 & 3 & Misaligned & 26\\ 
WASP-18 & 6 & 2007-09-16 & 2012-10-08 & 1849 & 2 & Control & 27\\ 
WASP-19 & 3 & 2008-05-29 & 2013-01-26 & 1702 & 4 & Control & 28,29\\ 
WASP-22 & 11 & 2008-08-26 & 2013-12-12 & 1934 & 3 & Control & 30\\ 
WASP-24 & 4 & 2009-01-01 & 2012-07-01 & 1277 & 4 & Control & 7,31\\ 
WASP-34 & 8 & 2009-12-01 & 2013-12-12 & 1472 & 2 & Control & 32\\ 
WASP-38 & 3 & 2010-03-30 & 2012-04-10 & 742 & 4 & Misaligned & 33\\ 
XO-2 & 9 & 2007-09-28 & 2013-12-14 & 2269 & 3 & Control & 10,34\tablenotemark{d}\\ 
XO-3 & 11 & 2006-09-27 & 2012-09-25 & 2190 & 5 & Misaligned & 35,36,37\\ 
XO-4 & 9 & 2007-12-21 & 2013-01-27 & 1864 & 3 & Misaligned & 38,39\\ 
XO-5 & 24 & 2007-03-27 & 2012-10-07 & 2021 & 2 & Control & 40\\ 
\enddata
\tablenotetext{a}{Total number of CPS radial velocity measurements excluding any in-transit data.}
\tablenotetext{b}{Number of independent data sets.  Although this usually refers to data taken by different telescopes, data obtained with HIRES before and after the CCD upgrade must also be treated as two separate data sets.}
\tablenotetext{c}{The misaligned sample consists of planets with either eccentric or misaligned orbits, while the control sample contains planets that appear to have circular and/or well-aligned orbits.}
\tablenotetext{d}{We exclude the WASP-10 FIES data from Christian et al. (2009) and the Burke et al. (2007) data for XO-2 as the error bars for these measurements were too large to justify the addition of another $\gamma$ parameter in our fit.}
\tablenotetext{e}{REFERENCES -
(1) \citet{hartman11b}; (2) \citet{kipping11}; (3) \citet{odonovan07}; (4) \citet{mandushev07}; (5) \citet{narita10a}; (6) \citet{cameron07}; (7) \citet{simpson11}; (8) \citet{albrecht11}; (9) \citet{pont11}; (10) \citet{husnoo12}; (11) \citet{pollacco08}; (12) \citet{simpson10}; (13) \citet{tripathi10}; (14) \citet{wilson08}; (15) \citet{hellier08}; (16) \citet{albrecht12a}; (17) \citet{queloz10}; (18) \citet{christian09}; (19) \citet{hebb09a}; (20) \citet{husnoo11}; (21) \citet{joshi09}; (22) \citet{johnson09}; (23) \citet{west09}; (24) \citet{lister09}; (25) \citet{brown12}; (26) \citet{anderson10}; (27) \citet{hellier09}; (28) \citet{hebb09b}; (29) \citet{hellier11}; (30) \citet{maxted10}; (31) \citet{street10}; (32) \citet{smalley11}; (33) \citet{barros11}; (34) \citet{narita11}; (35) \citet{johnskrull08}; (36) \citet{hebrard08}; (37) \citet{hirano11}; (38) \citet{mccullough08}; (39) \citet{narita10}; (40) \citet{burke08}
}
\end{deluxetable*}

\begin{deluxetable}{cccc}
\tablecaption{HIRES Radial Velocity Measurements\tablenotemark{a} \label{full_rv_table}}
\tablehead{
\colhead{BJD$_\text{TDB}$} & \colhead{RV (m s$^{-1}$)} & \colhead{Error (m s$^{-1}$)}  & \colhead{Star Name}
}
\startdata
2451552.07794 & 5.501 & 2.366 & GJ436 \\
2451706.86604 & -14.371 & 2.694 & GJ436 \\
2451983.01612 & 9.447 & 2.792 & GJ436 \\
2452064.87126 & 12.921 & 2.754 & GJ436 \\
2452308.08494 & 19.816 & 2.381 & GJ436 \\
2452333.03883 & -25.086 & 3.351 & GJ436 \\
2452334.05478 & 18.176 & 2.427 & GJ436 \\
2452363.03958 & 13.229 & 2.878 & GJ436 \\
2452711.8987 & -0.7 & 2.536 & GJ436 \\
2452804.87853 & 18.473 & 2.552 & GJ436 \\
\enddata
\tablenotetext{a}{The full table with measurements for all of the stars included in this study can be found in electronic format on the ApJ website, and is available from the authors upon request.}
\end{deluxetable}

In \S\ref{obs} we outline our target sample selection criteria and describe the acquisition of our radial velocity and adaptive optics data.  In \S\ref{analysis} we summarize our fits to the radial velocity data sets and the generation of contrast curves from our AO data.  In \S\ref{discussion} we discuss the implications of our results for the multiplicity fraction of hot Jupiters and constrain the masses and orbital separations of the companions.

\section{Observations}\label{obs}

\subsection{Sample Selection}

Our sample includes transiting planets with orbital periods between $0.7-11$ days and masses between $0.06-11$ M$_{\rm Jup}$ (i.e., planet with masses comparable to or larger than that of Neptune).  We focus our search on a sample of twenty seven systems where there is already evidence for multi-body dynamics, including planets with eccentric orbits or orbits that are significantly tilted with respect to the star's spin axis (see Fig.~\ref{ecc_vs_spin}).  We required that the planets in this sample have projected obliquities or eccentricities that differed from zero by more than $3\sigma$; for convenience we refer to this as the ``misaligned" sample, although we note that it also contains planets with eccentric orbits and obliquities consistent with zero.  We also include a control sample of twenty four planets that appear to have well-aligned and circular orbits (i.e., within $3\sigma$ of zero), where canonical disk migration models for isolated stars could plausibly explain the presence of the observed short period planet.  Because the stellar multiplicity rate increases for more massive stars, we select our control sample to have approximately the same distribution of stellar masses as our misaligned sample in order to avoid biasing our estimates of the companion frequencies (see Fig.~\ref{mass_hist} for the relative distribution and Table \ref{stellar_parameters} for a list of masses for individual systems).  A subset of the systems in our target list are known to exhibit radial velocity accelerations; in these cases, our data allow us to confirm and refine the properties of the long-period companion responsible for the trend.  

\subsection{Keck HIRES Radial Velocities}

We observed our target stars using the HIgh Resolution Echelle Spectrometer (HIRES) \citep{vogt94} on the 10 m Keck I telescope over a period of two years beginning in 2011; many of our targets also had existing HIRES observations taken prior to 2011 by other programs.  We used the standard HIRES setup and reduction pipeline employed by the California Planet Search (CPS) consortium \citep{wright04,howard09,johnson10}.  Observations were typically obtained with a slit width of $0.\arcsec 86$ with integration times optimized to obtain typical signal to noise ratios of 70 per pixel.  An iodine cell mounted in front of the spectrometer entrance slit provided a wavelength scale and instrumental profile for the observations \citep{marcy92,valenti95}.  We obtained a total of approximately 270 new radial velocity measurements for our target sample, with a minimum of four observations per target separated by at least six months.  We then combine our data with published radial velocities obtained using other telescopes to provide the strongest possible constraints on the presence of any long-term radial velocity accelerations.  We provide a summary of the radial velocity data utilized in this study in Table \ref{rv_obs}, as well as individual HIRES radial velocity measurements for each system in Table \ref{full_rv_table}.

\begin{deluxetable}{lllllll}
\tablecaption{Summary of Adaptive Optics Observations \label{ao_obs}}
\tablewidth{0pt}
\tablehead{
\colhead{Star} & \colhead{Obs. Date} & \colhead{Filter} & \colhead{Array\tablenotemark{a}} &\colhead{$T_{\rm int}$\tablenotemark{b}} & \colhead{N\tablenotemark{c}} 
}
\startdata
HAT-P-2 & UT 2012 May 29 & $K_p$ & 512 & 13.3 & 9 \\ 
HAT-P-4 & UT 2012 Feb 02 & $K_p$ & 1024 & 15 & 9 \\ 
HAT-P-7 & UT 2013 Jun 22 & $K_s$ & 1024 & 9 & 12  \\  
HAT-P-10 & UT 2012 Feb 02 & $K_p$ & 1024 & 10 & 9 \\ 
HAT-P-13 & UT 2012 Feb 02 & $K_p$ & 1024 & 9 & 9 \\ 
HAT-P-22 & UT 2012 Feb 02 & $K_p$ & 512 & 10 & 18 \\ 
HAT-P-29 & UT 2012 Feb 02 & $K_p$ & 1024 & 15 & 9 \\ 
HAT-P-32 & UT 2013 Mar 02 & $K_s$ & 1024 & 15 & 15 \\  
WASP-8 & UT 2012 Jul 27 & $K_p$ & 1024 & 9 & 30 \\ 
WASP-10 & UT 2012 Jul 4 & $K_p$ & 1024 & 20 & 9 \\ 
WASP-22 & UT 2012 Aug 26 & $K_p$ & 1024 & 10 & 9 \\ 
WASP-34 & UT 2012 Feb 02 & $K_p$ & 1024 & 10 & 18 \\ 
XO-2 & UT 2012 Feb 02 & $K_p$ & 1024 & 10 & 27 \\  
\enddata
\tablenotetext{a}{NIRC2 offers full ($1024\times1024$) array and subarray ($512\times512$) readout options; the smaller array allows for shorter minimum exposure times in order to avoid saturating on the brightest targets}.
\tablenotetext{b}{Total integration time in seconds for each image.}
\tablenotetext{c}{Total number of images acquired for each target.}
\end{deluxetable}

\begin{figure*}[ht]
\epsscale{1.1}
\plottwo{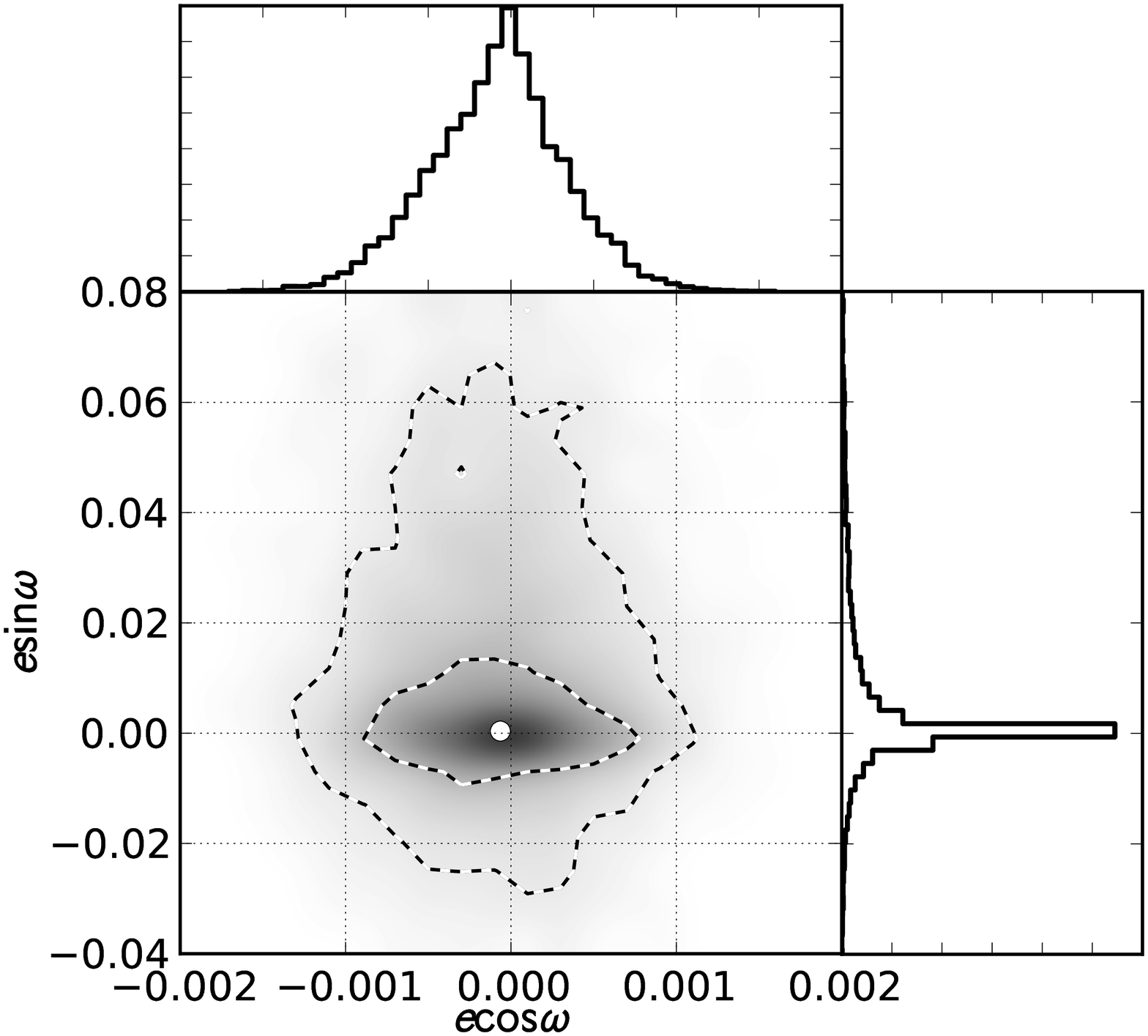}{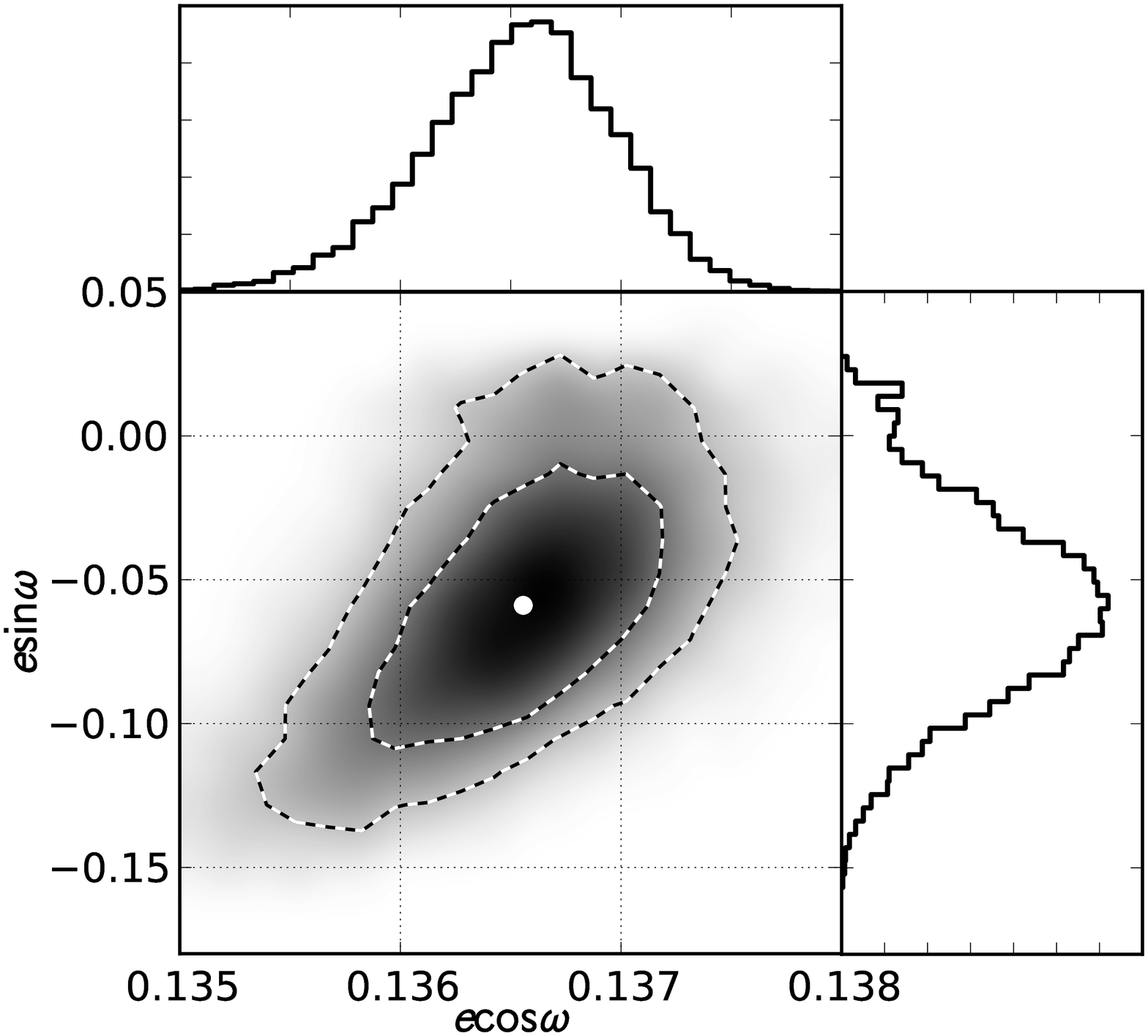}
\centering
\caption{
Two dimensional marginalized posterior distributions of $e\cos{\omega_{\star}}$ vs. $e\sin{\omega_{\star}}$ when secondary eclipse priors are applied. Left: Posterior distributions for HD149026, where the orbital phase of the secondary eclipse is very well known and the eccentricity is consistent with zero.  Right: Posterior distributions for GJ436. $e\cos{\omega_{\star}}$ and $e\sin{\omega_{\star}}$ become modestly correlated if the orbital phase of the secondary eclipse is very well known and the orbit is eccentric.}
\label{fig:distros}
\end{figure*} 

\subsection{NIRC2 AO Imaging}

In this paper we focus on images obtained for systems with detected radial velocity accelerations; we will present a complete analysis of our AO data set including companion detections in Paper II.  We obtained K band adaptive optics imaging \citep{wizinowich00} for each of our target stars using the NIRC2 instrument (Instrument PI: Keith Matthews) on Keck II in the narrow camera (10 mas pixel$^{-1}$) setting.  We used the full $1024\times1024$ pixel field of view for most of our target stars, with the exception of several of our brightest targets where we switched to a $512\times512$ pixel subarray in order to allow for shorter integration times and avoid saturation.  We utilized a standard three-point dither pattern \citep[e.g.,][]{bechter13} that maximizes our spatial coverage and allows for the removal of sky and instrumental backgrounds while avoiding the lower-left quadrant on the array, which has a higher read noise level.  We obtain our images in position angle mode, where the orientation of the image on the detector is kept constant as the telescope tracks, rather than using the angular differential imaging technique where the image is allowed to rotate on the detector and performing PSF subtraction.  This maximized the efficiency of our observations while still providing deep sensitivity to low-mass stellar companions \citep{crepp12}.

We flat-field our images and remove hot pixels by searching for $4\sigma$ outliers at a fixed pixel position, treating each nod position separately.   We calculate a median sky background using the off-nod positions and subtracting this median image from each of our science images at that nod position.  Each image is then interpolated by a factor of ten and the images are stacked using the point spread function of our target star in order to align the positions.  We create our final science images by taking the median flux at each pixel position in our stacked images.  A summary of the observations utilized in this analysis is provided in Table \ref{ao_obs}. 

\section{Analysis}\label{analysis}

\subsection{Radial Velocity Fits}
\label{rv_fits}

In order to detect and quantify the significance of long-term accelerations in the radial velocity data we performed a uniform analysis of all 51 systems with a Differential-evolution Markov Chain Monte Carlo \citep[DE-MCMC,][]{braak06} technique similar to that of \citet{fulton13}.  The DE-MCMC algorithm speeds convergence by downgrading the importance of pre-determining optimal step sizes for each parameter. DE-MCMC runs many chains in parallel (twice the number of free parameters) and uses the difference in parameter values from two random chains in order to establish the magnitude and direction of each step. This ensures that step sizes are optimized on-the-fly to achieve ideal acceptance rates ($\sim18\%$ for well-constrained fits) and high convergence rates. Step sizes for correlated parameters are automatically reduced in the direction orthogonal to the correlation which leads to fewer models being calculated in regions of parameter space that are highly disfavored by the data.

Our radial velocity model for each system was described by a minimum of 8 free parameters: period ($P$), time of mid-transit at a particular reference epoch (T$_{\rm mid}$), eccentricity ($e$), argument of periastron of the star's orbit ($\omega_\star$), velocity semi-amplitude (K), a relative radial velocity (RV) zero point ($\gamma$), slope ($\dot{\gamma}$), and RV ``jitter".  When required, we expanded on this baseline model by carrying out two-planet fits for systems where the outer companion's orbit exhibited significant curvature (HAT-P-17, WASP-8, WASP-34), and a three-planet fit for the HAT-P-13 system.  For some systems, data from multiple spectrographs were included and in these cases the relative RV zero-points ($\gamma$) were fit separately for each instrument.  GJ 436b has HIRES radial velocities obtained prior to the CCD upgrade, and we treat data before and after this upgrade as separate data sets with a different baseline normalization.   

RV ``jitter" may be dominated by instrumental effects as opposed to astrophysical noise and thus should not be expected to converge to the same value for different instruments \citep{isaacson10}. In order to prevent our fitted ``jitter" parameter from being driven to abnormally large values by particularly noisy datasets we first run a set of chains with a uniform jitter value for all datasets to obtain a best-fitting model. We then run the chains again, this time scaling the jitter value at each step by $\frac{\sigma_x}{\sigma_{\rm CPS}}$ where $\sigma_x$ is the RMS of the residuals to the best-fit model from the initial run for dataset $x$ and $\sigma_{CPS}$ is the RMS of the residuals of the best-fit model for the post-upgrade HIRES data. This ensures that the measurement errors are roughly equal to the RMS of the residuals to the final model for each individual dataset.  We also reject RV measurements from the CPS HIRES data with reported measurement errors that are greater than 10 times the median absolute deviation of all of the measurement errors for that particular star.  These measurements are typically derived from very low signal-to-noise spectra where the standard HIRES extraction routine does not produce optimal results, and contribute minimally to our fits.  This step generally results in the rejection of less than three outliers from each RV set. 

RV measurements taken during transits of the known planet were excluded from the fit.  For planets with high-cadence Rossiter measurements spanning several hours around the transit we take the error-weighted mean of the out-of-transit points and include this as a single measurement in our fits.  Because we add an additional jitter term, this effectively down-weights the contribution of these high-density data sets to our fit.  This conservative approach ensures that our best-fit solutions are not biased by the presence of short-term stellar variability that can cause trends in the RV measurements over several hour time scales \citep[e.g.][]{albrecht12a}.  

\clearpage
 \LongTables
\begin{landscape}
\begin{deluxetable}{l c c c c c}
\tablecaption{Priors Used in Radial Velocity Fits \label{rv_priors1}}
\tablewidth{530pt}
\tabletypesize{\tiny}
\tablehead{
\colhead{Planet} & \colhead{Period\tablenotemark{a}} & \colhead{T$_{\rm mid}$} & \colhead{Ephemeris reference} & \colhead{Secondary eclipse times} & \colhead{Secondary eclipse references} \\
\colhead{} & \colhead{days} & \colhead{\bjdtdb-2450000} & \colhead{} & \colhead{\bjdtdb-2450000} & \colhead{}
}
\startdata
GJ436b & 2.6438979 $\pm$ 3e-07 & 4865.083873 $\pm$ 4.2e-05 & \citet{knutson11} & \parbox[t]{3cm}{\centering4282.3336 $\pm$ 0.0016 \\  \vspace{0pt} 4628.6857 $\pm$ 0.0017 \\  \vspace{0pt} 4631.3288 $\pm$ 0.0021 \\  \vspace{0pt} 4633.9723 $\pm$ 0.0013 \\  \vspace{0pt} 4636.6169 $\pm$ 0.0021 \\  \vspace{0pt} 4660.4119 $\pm$ 0.0019 \\  \vspace{0pt} 4663.054 $\pm$ 0.004 \\  \vspace{0pt} 4858.7054 $\pm$ 0.0026 \\  \vspace{0pt} 4861.3467 $\pm$ 0.0015 \\  \vspace{0pt} 4863.9896 $\pm$ 0.0017 \\  \vspace{0pt} 4866.6362 $\pm$ 0.0023 \\ } & \citet{stevenson10} \\ 
HAT-P-2b & 5.6334729 $\pm$ 6.1e-06 & 5288.8498 $\pm$ 0.0006 & \citet{pal10} & \parbox[t]{3cm}{\centering5284.2966 $\pm$ 0.0014 \\  \vspace{0pt} 5751.8794 $\pm$ 0.0011 \\  \vspace{0pt} 4354.7757 $\pm$ 0.0022 \\ } & \citet{lewis13} \\ 
HAT-P-4b & 3.0565254 $\pm$ 1.2e-06 & 4245.8152 $\pm$ 0.0002 & \citet{sada12} & \parbox[t]{3cm}{\centering5298.7864 $\pm$ 0.0026 \\  \vspace{0pt} 5442.4437 $\pm$ 0.0032 \\ } & \citet{todorov13} \\ 
HAT-P-6b & 3.8530030 $\pm$ 1.2e-06 & 4035.67616 $\pm$ 0.00025 & \citet{todorov11} & \parbox[t]{3cm}{\centering5451.652 $\pm$ 0.004 \\  \vspace{0pt} 5459.3565 $\pm$ 0.0017 \\ } & \citet{todorov11} \\ 
HAT-P-7b & 2.204737 $\pm$ 1.7e-05 & 4954.357462 $\pm$ 5e-06 & \citet{morris13} & \parbox[t]{3cm}{\centering4768.0520 $\pm$ 0.0035 \\  \vspace{0pt} 4770.2640 $\pm$ 0.0039 \\ } & \citet{christiansen10} \\ 
HAT-P-8b & 3.0763402 $\pm$ 1.5e-06 & 4437.67657 $\pm$ 0.00034 & \citet{todorov11} & \parbox[t]{3cm}{\centering5211.3750 $\pm$ 0.0016 \\  \vspace{0pt} 5208.3010 $\pm$ 0.0024 \\ } & \citet{todorov11} \\ 
HAT-P-10b & 3.7224793 $\pm$ 7e-07 & 4759.68753 $\pm$ 0.00011 & \citet{sada12} &  & \\ 
HAT-P-11b & 4.8878056 $\pm$ 1.5e-06 & 4605.89123 $\pm$ 0.00013 & \citet{sada12} &  & \\ 
HAT-P-12b & 3.21305929 $\pm$ 3.4e-07 & 4187.85558 $\pm$ 0.00011 & \citet{todorov13} &  & \\ 
HAT-P-13b & 2.9162383 $\pm$ 2.2e-06 & 5176.53878 $\pm$ 0.00027 & \citet{southworth12} &  & \\ 
HAT-P-14b & 4.627669 $\pm$ 5e-06 & 5314.91866 $\pm$ 0.00066 & \citet{winn11} &  & \\ 
HAT-P-15b & 10.863502 $\pm$ 2.7e-05 & 4638.56094 $\pm$ 0.00048 & \citet{kovacs10} &  & \\ 
HAT-P-16b & 2.775960 $\pm$ 3e-06 & 5027.59369 $\pm$ 0.00031 & \citet{buchhave10} &  & \\ 
HAT-P-17b & 10.338523 $\pm$ 9e-06 & 4801.1702 $\pm$ 0.0003 & \citet{howard12} &  & \\ 
HAT-P-18b & 5.508023 $\pm$ 6e-06 & 4715.0224 $\pm$ 0.0002 & \citet{hartman11a} &  & \\ 
HAT-P-20b & 2.875317 $\pm$ 4e-06 & 5080.92737 $\pm$ 0.00021 & \citet{bakos11} &  & \\ 
HAT-P-22b & 3.212220 $\pm$ 9e-06 & 4930.22077 $\pm$ 0.00025 & \citet{bakos11} &  & \\ 
HAT-P-24b & 3.355240 $\pm$ 7e-06 & 5216.97743 $\pm$ 0.00028 & \citet{kipping10} &  & \\ 
HAT-P-26b & 4.234516 $\pm$ 1.5e-05 & 5304.65198 $\pm$ 0.00035 & \citet{hartman11b} &  & \\ 
HAT-P-29b & 5.723186 $\pm$ 4.9e-05 & 5197.57616 $\pm$ 0.00181 & \citet{buchhave11} &  & \\ 
HAT-P-30b & 2.810595 $\pm$ 5e-06 & 5456.46637 $\pm$ 0.00037 & \citet{johnson11} &  & \\ 
HAT-P-31b & 5.005425 $\pm$ 9.2e-05 & 4320.8865 $\pm$ 0.0052 & \citet{kipping11} &  & \\ 
HAT-P-32b & 2.150008 $\pm$ 1e-06 & 4420.44712 $\pm$ 9e-05 & \citet{hartman11c} &  & \\ 
HAT-P-33b & 3.474474 $\pm$ 1e-06 & 5110.92671 $\pm$ 0.00022 & \citet{hartman11c} &  & \\ 
HAT-P-34b & 5.452654 $\pm$ 1.6e-05 & 5431.59705 $\pm$ 0.00055 & \citet{bakos12} &  & \\ 
HD149026b & 2.8758916 $\pm$ 1.4e-06 & 4597.70712 $\pm$ 0.00016 & \citet{stevenson12} & \parbox[t]{3cm}{\centering4535.8768 $\pm$ 0.0012 \\  \vspace{0pt} 4596.268 $\pm$ 0.004 \\  \vspace{0pt} 4325.941 $\pm$ 0.011 \\  \vspace{0pt} 4633.65 $\pm$ 0.01 \\  \vspace{0pt} 4903.990 $\pm$ 0.013 \\  \vspace{0pt} 3606.964 $\pm$ 0.002 \\  \vspace{0pt} 4567.512 $\pm$ 0.004 \\  \vspace{0pt} 4599.132 $\pm$ 0.003 \\  \vspace{0pt} 4912.614 $\pm$ 0.002 \\  \vspace{0pt} 4317.311 $\pm$ 0.005 \\  \vspace{0pt} 4343.194 $\pm$ 0.005 \\ } & \citet{stevenson12} \\ 
\enddata
\tablenotetext{a}{For transiting planets the radial velocity data provide a relatively weak constraint on orbital period as compared to the transit ephemeris, so the period we derive from our fits to the RV data is indistinguishable from the input prior.}
\end{deluxetable}

\clearpage
 \LongTables
\begin{deluxetable}{l c c c c c}
\tablecaption{Priors Used in Radial Velocity Fits Continued\label{rv_priors2}}
\tablewidth{530pt}
\tabletypesize{\tiny}
\tablehead{
\colhead{Planet} & \colhead{Period\tablenotemark{a}} & \colhead{T$_{\rm mid}$} & \colhead{Ephemeris reference} & \colhead{Secondary eclipse times} & \colhead{Secondary eclipse references} \\
\colhead{} & \colhead{days} & \colhead{\bjdtdb-2450000} & \colhead{} & \colhead{\bjdtdb-2450000} & \colhead{}
}
\startdata
TrES-2b & 2.47061320 $\pm$ 2e-08 & 4955.7625504 $\pm$ 5.6e-06 & \citet{barclay12} & \parbox[t]{3cm}{\centering4994.0607 $\pm$ 0.0033 \\  \vspace{0pt} 4324.5227 $\pm$ 0.0026 \\  \vspace{0pt} 4070.04880 $\pm$ 0.00086 \\ } & \parbox[t]{5cm}{\centering \citet{croll10a} \\ \citet{odonovan10} \\}\\ 
TrES-3b & 1.3061854 $\pm$ 1e-07 & 4185.91289 $\pm$ 6e-05 & \citet{turner13} & \parbox[t]{3cm}{\centering4985.9550 $\pm$ 0.0014 \\  \vspace{0pt} 4668.545 $\pm$ 0.002 \\  \vspace{0pt} 4665.9350 $\pm$ 0.0027 \\  \vspace{0pt} 4668.550 $\pm$ 0.002 \\  \vspace{0pt} 4665.937 $\pm$ 0.002 \\ } & \parbox[t]{5cm}{\centering \citet{croll10b} \\ \citet{fressin10} \\}\\ 
TrES-4b & 3.5539303 $\pm$ 1.9e-06 & 4230.90574 $\pm$ 0.00043 & \citet{sada12} & \parbox[t]{3cm}{\centering4392.604 $\pm$ 0.011 \\  \vspace{0pt} 4396.1687 $\pm$ 0.0055 \\ } & \citet{knutson09} \\ 
WASP-1b & 2.5199425 $\pm$ 1.4e-06 & 3912.51531 $\pm$ 0.00032 & \citet{sada12} & \parbox[t]{3cm}{\centering$e\cos{\omega}$=0.0000 $\pm$ 0.0011 \\ } & \citet{wheatley10} \\ 
WASP-2b & 2.1522213 $\pm$ 4e-07 & 3991.51536 $\pm$ 0.00018 & \citet{sada12} & \parbox[t]{3cm}{\centering$e\cos{\omega}$=0.0000 $\pm$ 0.0013 \\ } & \citet{wheatley10} \\ 
WASP-3b & 1.8468332 $\pm$ 4e-07 & 4143.85193 $\pm$ 0.00017 & \citet{sada12} & \parbox[t]{3cm}{\centering5130.985 $\pm$ 0.002 \\  \vspace{0pt} 4728.3759 $\pm$ 0.0027 \\ } & Beerer et al., in prep \\ 
WASP-4b & 1.3382314 $\pm$ 3.2e-06 & 4697.798311 $\pm$ 4.6e-05 & \citet{nikolov12} & \parbox[t]{3cm}{\centering5174.87807 $\pm$ 0.00087 \\  \vspace{0pt} 5172.2018 $\pm$ 0.0013 \\ } & \citet{beerer11} \\ 
WASP-7b & 4.9546416 $\pm$ 3.5e-06 & 5446.6349 $\pm$ 0.0003 & \citet{albrecht12a} &  & \\ 
WASP-8b & 8.158715 $\pm$ 1.6e-05 & 4679.33393 $\pm$ 0.00047 & \citet{queloz10} & \parbox[t]{3cm}{\centering5401.4989 $\pm$ 0.0028 \\  \vspace{0pt} 4822.2308 $\pm$ 0.0031 \\  \vspace{0pt} 4814.0739 $\pm$ 0.0033 \\  \vspace{0pt} 5409.6663 $\pm$ 0.0023 \\ } & \citet{cubillos13} \\ 
WASP-10b & 3.0927293 $\pm$ 3.2e-06 & 4664.038089 $\pm$ 4.8e-05 & \citet{barros13} &  & \\ 
WASP-12b & 1.0914224 $\pm$ 3e-07 & 4508.97683 $\pm$ 0.00019 & \citet{sada12} & \parbox[t]{3cm}{\centering4773.6480 $\pm$ 0.0006 \\  \vspace{0pt} 4769.2818 $\pm$ 0.0008 \\ } & \citet{campo11} \\ 
WASP-14b & 2.2437704 $\pm$ 2.8e-06 & 4963.93752 $\pm$ 0.00025 & \citet{johnson09} & \parbox[t]{3cm}{\centering5274.6617 $\pm$ 0.0007 \\  \vspace{0pt} 4908.9295 $\pm$ 0.0011 \\ } & \citet{blecic12} \\ 
WASP-15b & 3.7520656 $\pm$ 2.8e-06 & 4584.69823 $\pm$ 0.00029 & \citet{west09} &  & \\ 
WASP-16b & 3.11860 $\pm$ 1e-05 & 4584.42951 $\pm$ 0.00029 & \citet{lister09} &  & \\ 
WASP-17b & 3.7354845 $\pm$ 1.9e-06 & 4592.8015 $\pm$ 0.0005 & \citet{southworth12c} &  & \\ 
WASP-18b & 0.9414523 $\pm$ 3e-07 & 5265.5525 $\pm$ 0.0001 & \citet{maxted13} & \parbox[t]{3cm}{\centering4820.7159 $\pm$ 0.0007 \\  \vspace{0pt} 4824.4807 $\pm$ 0.0006 \\ } & \citet{nymeyer10} \\ 
WASP-19b & 0.78883942 $\pm$ 3.3e-07 & 4775.33754 $\pm$ 0.00018 & \citet{tregloan12} & \parbox[t]{3cm}{\centering$\Phi_{s}$=0.50005 $\pm$ 0.00048 \\ } & \citet{anderson13} \\ 
WASP-22b & 3.5327313 $\pm$ 5.8e-06 & 5497.40042 $\pm$ 0.00025 & \citet{anderson11} &  & \\ 
WASP-24b & 2.3412162 $\pm$ 1.4e-06 & 5081.3803 $\pm$ 0.0001 & \citet{sada12} & \parbox[t]{3cm}{\centering$\Phi_{s}$=0.50027 $\pm$ 0.00056 \\ } & \citet{smith12} \\ 
WASP-34b & 4.3176782 $\pm$ 4.5e-06 & 4647.55434 $\pm$ 0.00064 & \citet{smalley11} &  & \\ 
WASP-38b & 6.871815 $\pm$ 4.4e-05 & 5335.92128 $\pm$ 0.00074 & \citet{barros11} &  & \\ 
XO-2b & 2.61586178 $\pm$ 7.5e-07 & 5981.46035 $\pm$ 0.00013 & \citet{sing12} & \parbox[t]{3cm}{\centering4421.104 $\pm$ 0.021 \\  \vspace{0pt} 4423.723 $\pm$ 0.018 \\ } & \citet{machalek09} \\ 
XO-3b & 3.1915289 $\pm$ 3.2e-06 & 4864.7668 $\pm$ 0.0004 & \citet{winn09a} & \parbox[t]{3cm}{\centering4908.402 $\pm$ 0.017 \\  \vspace{0pt} 4943.50 $\pm$ 0.02 \\ } & \citet{machalek10} \\ 
XO-4b & 4.1250823 $\pm$ 3.9e-06 & 4485.93306 $\pm$ 0.00036 & \citet{todorov11} & \parbox[t]{3cm}{\centering5181.0175 $\pm$ 0.0062 \\  \vspace{0pt} 5172.7595 $\pm$ 0.0016 \\ } & \citet{todorov11} \\ 
XO-5b & 4.1877545 $\pm$ 1.6e-06 & 4485.66875 $\pm$ 0.00028 & \citet{sada12} &  & \\ 
\enddata
\tablenotetext{a}{For transiting planets the radial velocity data provide a relatively weak constraint on orbital period as compared to the transit ephemeris, so the period we derive from our fits to the RV data is indistinguishable from the input prior.}
\end{deluxetable}
\clearpage
\end{landscape}

\clearpage
 \LongTables
 \begin{landscape}
\begin{deluxetable}{l l l l l l l l l l l l}
\tablecaption{Results from Radial Velocity Fits \label{rv_params1}}
\tablewidth{610pt}
\tabletypesize{\tiny}
\tablehead{
\colhead{Planet} & \colhead{$e$} & \colhead{$\omega_{\star}$} & \colhead{$e\cos{\omega_{\star}}$} & \colhead{$e\sin{\omega_{\star}}$} & \colhead{$K$} & \colhead{$\dot{\gamma}$\tablenotemark{a}} & \colhead{jitter} & \colhead{M$_p$} & \colhead{Spin-Orbit $\lambda$} & \colhead{Sample\tablenotemark{b}} & \colhead{Ref.}\\
\colhead{} & \colhead{} & \colhead{(degrees)} & \colhead{} & \colhead{}& \colhead{(\ms)} & \colhead{(\ms day$^{-1}$)} & \colhead{(\ms)} & \colhead{M$_{\rm Jup}$} & \colhead{(degrees)} & \colhead{} & \colhead{}
}
\startdata
GJ436b &  0.1495 $^{+0.016}_{-0.0097}$  &  336 $^{+12}_{-11}$\tablenotemark{c}  &  0.13654 $^{+0.0004}_{-0.00047}$  &  -0.061 $^{+0.032}_{-0.033}$  &  17.01 $\pm 0.54$\tablenotemark{c}  &  -0.00137 $\pm 0.00061$  &  3.78 $^{+0.32}_{-0.29}$  & 0.0682 $\pm 0.0025$ & & Misaligned & 1\\ 
HAT-P-2b &  0.5079 $^{+0.00093}_{-0.00079}$  &  186.2 $\pm 1.1$  &  -0.50493 $^{+0.00043}_{-0.00042}$  &  -0.0546 $\pm 0.0099$  &  929 $\pm 11$  &  \textbf{-0.0938 $^{+0.0067}_{-0.0069}$}  &  31.7 $^{+4.3}_{-3.5}$  & 9.99 $\pm 0.23$ & $9\pm10$ & Misaligned & 3,4 \\ 
HAT-P-4b &  0.004 $^{+0.017}_{-0.0031}$  &  157 $^{+110}_{-67}$  &  -0.00055 $^{+0.00084}_{-0.0011}$  &  0.0001 $^{+0.011}_{-0.0075}$  &  77 $\pm 3$\tablenotemark{d} &  \textbf{0.0219 $\pm 0.0035$}  &  9.9 $^{+2.1}_{-1.6}$  & 0.639 $^{+0.043}_{-0.042}$ & $-4.9\pm11.9$ & Control & 1,5\\ 
HAT-P-6b &  0.023 $^{+0.022}_{-0.02}$  &  94.1 $^{+37.0}_{-2.3}$  &  -0.00159 $^{+0.00066}_{-0.00067}$  &  0.023 $\pm 0.022$  &  120.8 $\pm 2.4$  &  0.0041 $\pm 0.0019$  &  6.6 $^{+1.9}_{-1.6}$  & 1.107 $\pm 0.041$ & $165\pm6$ & Misaligned & 1,3 \\ 
HAT-P-7b &  0.0055 $^{+0.007}_{-0.0033}$  &  204 $^{+53}_{-89}$  &  -0.003 $\pm 0.002$  &  -0.0008 $^{+0.0053}_{-0.0086}$  &  214.3 $^{+2.6}_{-2.5}$  &  \textbf{0.0646 $^{+0.004}_{-0.0038}$}  &  12.8 $^{+1.7}_{-1.4}$  & 1.697 $^{+0.027}_{-0.026}$ & $155\pm37$ & Misaligned & 3,6\\ 
HAT-P-8b &  0.0029 $^{+0.016}_{-0.0024}$  &  116 $^{+150}_{-35}$  &  -7e-05 $^{+0.0007}_{-0.00069}$  &  0.0003 $^{+0.014}_{-0.0035}$  &  162.0 $^{+4.4}_{-3.9}$\tablenotemark{d}  &  -0.0003 $^{+0.0034}_{-0.004}$  &  9.1 $^{+3.5}_{-2.3}$  & 1.304 $^{+0.065}_{-0.064}$ & $-17^{+9.2}_{-11.5}$ & Control & 7,8\\ 
HAT-P-10b &  0.028 $^{+0.029}_{-0.02}$  &  146 $^{+95}_{-52}$  &  -0.011 $^{+0.015}_{-0.026}$  &  0.008 $^{+0.031}_{-0.017}$  &  75.7 $^{+2.7}_{-2.6}$  &  \textbf{-0.014 $^{+0.0032}_{-0.0031}$}  &  6.1 $^{+2.1}_{-1.4}$  & 0.509 $\pm 0.023$ & & Control & 9\\ 
HAT-P-11b &  0.232 $^{+0.054}_{-0.053}$  &  7 $^{+24}_{-25}$  &  0.213 $^{+0.049}_{-0.053}$  &  0.028 $^{+0.097}_{-0.092}$  &  10.2 $^{+1.1}_{-1.2}$  &  \textbf{0.0094 $\pm 0.0016$}\tablenotemark{e}  &  5.95 $^{+0.58}_{-0.52}$  & 0.0756 $\pm 0.0087$ & $103^{+26}_{-10}$ & Misaligned & 10,11\\ 
HAT-P-12b &  0.026 $^{+0.026}_{-0.018}$  &  97 $^{+220}_{-64}$  &  0.012 $^{+0.021}_{-0.014}$  &  0.004 $^{+0.031}_{-0.019}$  &  35.4 $\pm 1.6$  &  -0.0004 $^{+0.0018}_{-0.0019}$  &  4.23 $^{+1.1}_{-0.92}$  & 0.2089 $^{+0.01}_{-0.0097}$ & & Control & 12\\ 
HAT-P-13b &  0.0133 $^{+0.0047}_{-0.0044}$  &  197 $^{+32}_{-37}$  &  -0.0107 $^{+0.0039}_{-0.0041}$  &  -0.0032 $^{+0.0073}_{-0.0077}$  &  105.87 $\pm 0.78$  &  \textbf{0.0528 $^{+0.0013}_{-0.0014}$}  &  4.53 $^{+0.54}_{-0.46}$  & 0.899 $^{+0.03}_{-0.029}$ & $1.9\pm8.6$ & Misaligned & 13\\ 
HAT-P-14b &  0.115 $^{+0.015}_{-0.016}$  &  98.8 $^{+5.4}_{-5.2}$  &  -0.02 $\pm 0.01$  &  0.113 $^{+0.015}_{-0.016}$  &  222.4 $\pm 3.7$  &  -0.0138 $^{+0.0062}_{-0.0061}$  &  9.1 $^{+3.2}_{-2.2}$  & 2.316 $\pm 0.072$ & $-170.9\pm5.1$ & Misaligned & 14,15\\ 
HAT-P-15b &  0.208 $^{+0.026}_{-0.025}$  &  261.8 $^{+2.2}_{-2.4}$  &  -0.0296 $^{+0.0076}_{-0.0077}$  &  -0.206 $^{+0.025}_{-0.026}$  &  185.8 $^{+5.2}_{-5.1}$  &  0.0131 $^{+0.0056}_{-0.0059}$  &  17.3 $^{+3.0}_{-2.3}$  & 2.043 $^{+0.082}_{-0.08}$ & & Misaligned & 16\\ 
HAT-P-16b &  0.0423 $^{+0.01}_{-0.0077}$  &  215 $^{+14}_{-21}$  &  -0.0342 $^{+0.0045}_{-0.0041}$  &  -0.023 $\pm 0.015$  &  534.1 $^{+6.3}_{-6.2}$  &  0.005 $^{+0.011}_{-0.01}$  &  10.6 $^{+5.6}_{-3.1}$  & 4.22 $\pm 0.11$ & $-10\pm16$ &Misaligned & 8,17\\ 
HAT-P-17b &  0.342 $\pm 0.0039$  &  199.1 $\pm 1.7$  &  -0.3229 $^{+0.004}_{-0.0041}$  &  -0.11 $\pm 0.01$  &  59.98 $\pm 0.79$  &  \textbf{$\equiv$ 0.0 $\pm 0.0$}\tablenotemark{f}  &  1.5 $\pm 0.45$  & 0.58 $\pm 0.019$  & $19^{+14}_{-16}$ & Misaligned & 18,19\\ 
HAT-P-18b &  0.106 $^{+0.15}_{-0.084}$  &  12 $^{+22}_{-21}$  &  0.095 $^{+0.13}_{-0.082}$  &  0.008 $^{+0.099}_{-0.018}$  &  25.4 $^{+5.8}_{-4.5}$  &  0.0004 $^{+0.0085}_{-0.0077}$  &  17.5 $^{+2.5}_{-2.4}$  & 0.183 $^{+0.034}_{-0.032}$ & & Control & 20\\ 
HAT-P-20b &  0.0158 $^{+0.0041}_{-0.0036}$  &  327 $^{+19}_{-13}$  &  0.013 $^{+0.0023}_{-0.0025}$  &  -0.0084 $^{+0.0053}_{-0.0052}$  &  1245.4 $^{+6.1}_{-6.3}$  &  -0.0141 $^{+0.0073}_{-0.0078}$  &  14.3 $^{+4.5}_{-3.2}$  & 7.24 $\pm 0.18$ & & Misaligned & 21\\ 
HAT-P-22b &  0.0064 $^{+0.008}_{-0.0046}$  &  116 $^{+180}_{-57}$  &  0.0002 $^{+0.0045}_{-0.004}$  &  0.0018 $^{+0.0099}_{-0.0048}$  &  314.4 $\pm 3.2$  &  \textbf{-0.0147 $^{+0.0043}_{-0.0045}$}  &  9.7 $^{+2.2}_{-1.6}$  & 2.157 $\pm 0.059$ & & Control & 21\\ 
HAT-P-24b &  0.033 $^{+0.027}_{-0.021}$  &  182 $\pm 63$  &  -0.02 $^{+0.019}_{-0.023}$  &  -0.0004 $^{+0.027}_{-0.028}$  &  86.5 $\pm 3.6$  &  -0.0099 $^{+0.0072}_{-0.0071}$  &  10.8 $^{+3.2}_{-2.6}$  & 0.715 $\pm 0.035$ & $20\pm16$ & Control & 3,22\\ 
HAT-P-26b &  0.14 $^{+0.12}_{-0.08}$  &  46 $^{+33}_{-71}$  &  0.075 $^{+0.062}_{-0.065}$  &  0.074 $^{+0.15}_{-0.097}$  &  8.57 $^{+0.99}_{-0.97}$  &  0.002 $\pm 0.002$  &  3.0 $^{+0.74}_{-0.62}$  & 0.0595 $^{+0.0072}_{-0.0071}$ & & Control & 23\\ 
HAT-P-29b &  0.061 $^{+0.044}_{-0.036}$  &  211 $^{+39}_{-65}$  &  -0.04 $^{+0.034}_{-0.031}$  &  -0.02 $^{+0.038}_{-0.057}$  &  77.6 $^{+4.5}_{-4.6}$  & \textbf{0.0498 $^{+0.0092}_{-0.01}$}  &  10.8 $^{+4.0}_{-2.6}$  & 0.773 $^{+0.052}_{-0.051}$ & & Control & 24\\ 
HAT-P-30b &  0.02 $^{+0.022}_{-0.014}$  &  114 $^{+200}_{-77}$  &  0.008 $^{+0.016}_{-0.01}$  &  0.002 $^{+0.024}_{-0.016}$  &  89.8 $^{+2.7}_{-2.8}$  &  0.0112 $^{+0.0068}_{-0.0071}$  &  7.7 $^{+2.0}_{-1.4}$  & 0.726 $\pm 0.027$ & $73.5\pm9.0$ & Misaligned & 25\\ 
HAT-P-31b &  0.2419 $^{+0.0099}_{-0.0097}$  &  276.2 $\pm 1.8$  &  0.0262 $^{+0.0076}_{-0.0075}$  &  -0.2404 $^{+0.0097}_{-0.0099}$  &  231.6 $^{+2.5}_{-2.6}$  &  0.0054 $^{+0.0072}_{-0.007}$  &  6.6 $^{+1.7}_{-1.3}$  & 2.227 $^{+0.089}_{-0.09}$ & & Misaligned & 26\\ 
HAT-P-32b &  0.2 $^{+0.19}_{-0.13}$  &  58 $^{+28}_{-53}$  &  0.076 $^{+0.11}_{-0.079}$  &  0.15 $^{+0.19}_{-0.15}$  &  112 $^{+20}_{-21}$  &  \textbf{-0.097 $\pm 0.023$}  &  64 $^{+11}_{-9}$  & 0.79 $\pm 0.15$ & $85\pm1.5$ & Misaligned & 3,27\\ 
HAT-P-33b &  0.13 $^{+0.19}_{-0.1}$  &  15 $\pm 22$  &  0.114 $^{+0.16}_{-0.097}$  &  0.015 $^{+0.13}_{-0.023}$  &  72 $^{+19}_{-16}$  &  -0.021 $^{+0.02}_{-0.023}$  &  53.5 $^{+12.0}_{-8.1}$  & 0.65 $\pm 0.14$ & & Control & 27\\ 
HAT-P-34b &  0.411 $^{+0.029}_{-0.028}$  &  17.8 $\pm 7.2$  &  0.388 $^{+0.027}_{-0.028}$  &  0.124 $^{+0.054}_{-0.051}$  &  364 $^{+24}_{-25}$  &  0.071 $^{+0.048}_{-0.05}$  &  52 $^{+14}_{-10}$  & 3.93 $\pm 0.28$\tablenotemark{c} & $0\pm14$ & Misaligned & 3,28\\ 
HD149026b &  0.0028 $^{+0.019}_{-0.0024}$  &  100 $^{+170}_{-11}$  &  -5e-05 $^{+0.00036}_{-0.00045}$  &  0.0005 $^{+0.021}_{-0.0025}$  &  37.9 $^{+1.4}_{-1.3}$\tablenotemark{d}  &  -0.00098 $^{+0.00099}_{-0.00089}$  &  5.13 $^{+0.85}_{-0.62}$  & 0.324 $\pm 0.011$\tablenotemark{d} & $12\pm7$ & Control & 2,3\\ 
TrES-2b &  0.0036 $^{+0.015}_{-0.0027}$  &  24 $^{+63}_{-110}$  &  0.00076 $^{+0.00053}_{-0.00052}$  &  0.0002 $^{+0.011}_{-0.0061}$  &  180.1 $^{+5.7}_{-5.6}$  &  -0.0041 $^{+0.006}_{-0.0059}$  &  17.9 $^{+4.0}_{-2.8}$  & 1.157 $^{+0.055}_{-0.056}$ & $-9\pm12$ & Control & 1,29\\ 
TrES-3b &  0.17 $^{+0.032}_{-0.031}$  &  270.5 $^{+0.38}_{-0.32}$  &  0.00151 $^{+0.001}_{-0.00098}$  &  -0.17 $^{+0.031}_{-0.032}$  &  312 $^{+13}_{-12}$\tablenotemark{d}  &  0.08 $^{+0.053}_{-0.054}$  &  104 $^{+60}_{-31}$  & 1.615 $^{+0.079}_{-0.077}$\tablenotemark{d} & & Misaligned & 30\\ 
TrES-4b &  0.015 $^{+0.076}_{-0.012}$  &  80.6 $^{+9.5}_{-160.0}$  &  0.0012 $^{+0.0022}_{-0.0018}$  &  0.005 $^{+0.082}_{-0.011}$  &  84 $\pm 10$  &  0.015 $\pm 0.012$  &  16.5 $^{+5.7}_{-3.9}$  & 0.843 $^{+0.098}_{-0.089}$ & $6.3\pm4.7$ & Control & 31,32\\ 
WASP-1b &  .0082 $^{+0.026}_{-0.0072}$  &  91.1 $^{+170.0}_{-6.3}$  &  3e-05 $\pm 0.001$  &  0.0068 $^{+0.028}_{-0.0072}$  &  119.6 $^{+3.1}_{-3.3}$  &  0.0029 $^{+0.0057}_{-0.0056}$  &  8.6 $^{+3.0}_{-2.3}$  & 0.79 $\pm 0.033$ & $-59\pm99$ & Control & 33,34\\ 
WASP-2b &  0.0054 $^{+0.009}_{-0.0044}$  &  267 $^{+11}_{-86}$  &  -0.0001 $^{+0.001}_{-0.0011}$  &  -0.0051 $^{+0.0051}_{-0.0092}$  &  156.7 $^{+1.2}_{-1.3}$  &  0.0062 $\pm 0.0092$  &  2.75 $^{+0.75}_{-0.63}$  & 0.918 $^{+0.027}_{-0.028}$ & $153^{+11}_{-15}$ & Misaligned & 35\\ 
WASP-3b &  0.0066 $^{+0.016}_{-0.0052}$  &  79 $^{+10}_{-130}$  &  0.0011 $^{+0.0014}_{-0.0012}$  &  0.0052 $^{+0.017}_{-0.0059}$  &  284.0 $^{+5.7}_{-6.0}$  &  -0.0126 $^{+0.0091}_{-0.0085}$  &  15.5 $^{+4.5}_{-3.6}$  & 1.944 $^{+0.04}_{-0.042}$ & $3.3^{2.5}_{-4.4}$ & Control & 36,37\\ 
WASP-4b &  0.0034 $^{+0.0074}_{-0.0026}$  &  288 $^{+140}_{-21}$  &  0.0006 $^{+0.0014}_{-0.0011}$  &  -0.0019 $^{+0.0027}_{-0.0087}$  &  234.6 $^{+2.2}_{-2.3}$  &  -0.0099 $^{+0.0052}_{-0.0054}$  &  1.91 $^{+0.29}_{-0.24}$  & 1.159 $^{+0.063}_{-0.064}$ & $-1^{+14}_{-12}$ & Control & 35,38\\ 
\enddata
\tablenotetext{a}{Systems with accelerations that differ from zero by more than $3\sigma$ are marked in bold.}
\tablenotetext{b}{The misaligned sample consists of planets with either eccentric or misaligned orbits, while the control sample contains planets that appear to have circular and/or well-aligned orbits.}
\tablenotetext{c}{Our preferred value for this parameter differs from the one in the published literature; this is likely related to our treatment of the stellar jitter and Rossiter data (see \S\ref{rv_fits}).}
\tablenotetext{d}{Our value for this parameter differs from the literature, but previous fits were calculated assuming a circular orbit.}
\tablenotetext{e}{The radial velocity acceleration in this system appears to be correlated with the stellar activity, and we therefore conclude that this is probably not the result of an additional companion in this system; see \S\ref{trend_discussion} for more details.}
\tablenotetext{f}{Because the acceleration in the HAT-P-17 system has some curvature, we fit it with a two-planet solution where the linear trend slope term is fixed to zero (see \citet{fulton13} for the full solution).}
\end{deluxetable}

\clearpage
 \LongTables
\begin{deluxetable}{l l l l l l l l l l l l}
\tablecaption{Results from Radial Velocity Fits Continued \label{rv_params2}}
\tablewidth{610pt}
\tabletypesize{\tiny}
\tablehead{
\colhead{Planet} & \colhead{$e$} & \colhead{$\omega_{\star}$} & \colhead{$e\cos{\omega_{\star}}$} & \colhead{$e\sin{\omega_{\star}}$} & \colhead{$K$} & \colhead{$\dot{\gamma}$\tablenotemark{a}} & \colhead{jitter} & \colhead{M$_p$} & \colhead{Spin-Orbit $\lambda$} & \colhead{Sample\tablenotemark{b}} & \colhead{Ref.}\\
\colhead{} & \colhead{} & \colhead{(degrees)} & \colhead{} & \colhead{}& \colhead{(\ms)} & \colhead{(\ms day$^{-1}$)} & \colhead{(\ms)} & \colhead{M$_{\rm Jup}$} & \colhead{(degrees)} & \colhead{} & \colhead{}
}
\startdata 
WASP-7b &  0.034 $^{+0.045}_{-0.024}$  &  109 $^{+170}_{-55}$  &  0.003 $^{+0.026}_{-0.02}$  &  0.011 $^{+0.055}_{-0.025}$  &  111.3 $^{+7.3}_{-7.5}$  &  0.03 $\pm 0.04$  &  34.6 $^{+4.5}_{-3.8}$  & 1.131 $^{+0.092}_{-0.089}$ & $86\pm8$ & Misaligned &  3, 39\\ 
WASP-8b &  0.3044 $^{+0.0039}_{-0.004}$  &  274.215 $^{+0.084}_{-0.082}$  &  0.02237 $^{+0.00032}_{-0.00031}$  &  -0.304 $\pm 0.004$  &  221.1 $\pm 1.2$  &  \textbf{$\equiv$ 0.0 $\pm 0.0$}\tablenotemark{c}  &  2.91 $^{+0.4}_{-0.34}$  & 2.24 $^{+0.11}_{-0.12}$ & $-123^{+3.4}_{-4.4}$ & Misaligned & 40\\ 
WASP-10b &  0.0473 $^{+0.0034}_{-0.0029}$  &  165.6 $^{+9.6}_{-8.6}$  &  -0.0454 $^{+0.0024}_{-0.0023}$  &  0.0118 $^{+0.0076}_{-0.008}$  &  568.8 $^{+7.0}_{-6.7}$  &  \textbf{-0.048 $^{+0.013}_{-0.012}$}  &  5.4 $^{+1.8}_{-1.3}$  & 3.37 $\pm 0.11$ & & Misaligned & 41\\ 
WASP-12b &  0.037 $^{+0.014}_{-0.015}$  &  272.7 $^{+2.4}_{-1.3}$  &  0.00171 $^{+0.00073}_{-0.00075}$  &  -0.037 $^{+0.015}_{-0.014}$  &  220.2 $\pm 3.1$  &  -0.0009 $^{+0.0097}_{-0.0093}$  &  19.5 $^{+2.7}_{-2.3}$  & 1.39 $\pm 0.13$ & $59^{+15}_{-20}$ & Misaligned & 3,42\\ 
WASP-14b &  0.0822 $^{+0.003}_{-0.0032}$\tablenotemark{d}  &  251.67 $^{+0.64}_{-0.75}$\tablenotemark{d} &  -0.02591 $^{+0.00049}_{-0.00046}$  &  -0.078 $^{+0.0034}_{-0.0032}$  &  987.2 $^{+1.7}_{-1.8}$  &  0.0062 $^{+0.0044}_{-0.0041}$  &  5.69 $^{+1.1}_{-0.85}$  & 7.8 $^{+0.45}_{-0.47}$ & $-33.1\pm7.4$ & Misaligned & 43,44\\ 
WASP-15b &  0.038 $^{+0.043}_{-0.026}$  &  240 $^{+79}_{-200}$  &  0.015 $^{+0.027}_{-0.018}$  &  -0.002 $^{+0.038}_{-0.044}$  &  61.8 $^{+4.6}_{-4.5}$  &  0.052 $^{+0.047}_{-0.044}$  &  4.4 $^{+2.5}_{-2.3}$  & 0.566 $\pm 0.045$ & $-139.6^{4.3}_{-4.2}$ & Misaligned & 35\\ 
WASP-16b &  0.015 $^{+0.012}_{-0.011}$  &  97 $^{+44}_{-20}$  &  -0.0009 $^{+0.004}_{-0.0048}$  &  0.014 $\pm 0.013$  &  118.9 $\pm 1.6$  &  0.0056 $^{+0.0071}_{-0.0072}$  &  2.34 $\pm 0.59$  & 0.846 $\pm 0.033$& $11^{+26}_{-19}$ & Control & 3,45\\ 
WASP-17b &  0.039 $^{+0.05}_{-0.027}$  &  179 $\pm 120$  &  0.006 $^{+0.031}_{-0.021}$  &  0.0001 $^{+0.047}_{-0.044}$  &  58.8 $^{+4.4}_{-4.7}$  &  0.0002 $^{+0.024}_{-0.026}$  &  11.7 $^{+5.0}_{-4.4}$  & 0.529 $^{+0.047}_{-0.048}$ & $-148.7^{+7.7}_{-6.7}$ & Misaligned & 35,46\\ 
WASP-18b &  0.0068 $^{+0.0025}_{-0.0027}$  &  261.1 $^{+5.3}_{-7.4}$  &  -0.00104 $^{+0.00065}_{-0.00067}$  &  -0.0067 $^{+0.0028}_{-0.0025}$  &  1816.6 $^{+6.1}_{-6.3}$  &  -0.003 $^{+0.0072}_{-0.0077}$  &  5.1 $^{+2.6}_{-1.9}$  & 10.47 $^{+0.49}_{-0.5}$ & $13\pm7$ & Control & 3,35\\ 
WASP-19b &  0.0024 $^{+0.0094}_{-0.0019}$  &  260 $^{+15}_{-170}$  &  -7e-05 $^{+0.00062}_{-0.00066}$  &  -0.0007 $^{+0.0019}_{-0.01}$  &  254.0 $^{+3.4}_{-3.3}$  &  0.065 $\pm 0.034$  &  17.8 $^{+3.2}_{-2.7}$  & 1.123 $\pm 0.036$ & $1.0\pm1.2$ & Control & 47,48\\ 
WASP-22b &  0.0108 $^{+0.014}_{-0.0076}$  &  114 $^{+160}_{-56}$  &  0.0005 $^{+0.0079}_{-0.0066}$  &  0.003 $^{+0.018}_{-0.008}$  &  70.9 $^{+1.5}_{-1.6}$  &  \textbf{0.0583 $^{+0.0078}_{-0.0074}$}  &  7.2 $^{+1.7}_{-1.4}$  & 0.569 $^{+0.016}_{-0.015}$ & $22\pm16$ & Control & 49,50\\ 
WASP-24b &  0.0033 $^{+0.012}_{-0.0026}$  &  70 $^{+20}_{-150}$  &  0.0005 $^{+0.00086}_{-0.0007}$  &  0.0011 $^{+0.014}_{-0.0027}$  &  152.0 $\pm 3.2$  &  -0.062 $\pm 0.051$  &  3.65 $^{+0.89}_{-0.8}$  & 1.119 $\pm 0.029$ & $-4.7\pm4$ & Control & 7,33\\ 
WASP-34b &  .0109 $^{+0.015}_{-0.0078}$  &  215 $^{+77}_{-140}$  &  -0.0001 $^{+0.0068}_{-0.0071}$  &  -0.001 $^{+0.011}_{-0.017}$  &  71.1 $^{+1.6}_{-1.7}$  &  \textbf{$\equiv$ 0.0 $\pm 0.0$\tablenotemark{c}}  &  3.2 $^{+0.72}_{-0.6}$  & 0.57 $\pm 0.03$  & & Control & 51\\ 
WASP-38b &  0.0329 $^{+0.01}_{-0.0086}$  &  17 $^{+27}_{-33}$  &  0.0284 $^{+0.0076}_{-0.0078}$  &  0.008 $^{+0.018}_{-0.016}$  &  252.1 $^{+4.4}_{-4.3}$  &  -0.074 $\pm 0.058$  &  11.9 $^{+2.4}_{-1.9}$  & 2.705 $\pm 0.076$ & $7.5^{+4.7}_{-6.1}$ & Misaligned & 52\\ 
XO-2b &  0.028 $^{+0.038}_{-0.022}$  &  261 $^{+11}_{-71}$  &  -0.0037 $^{+0.0052}_{-0.0063}$  &  -0.027 $^{+0.027}_{-0.039}$  &  93.9 $^{+2.1}_{-2.2}$  &  \textbf{0.0126 $^{+0.0039}_{-0.0036}$}  &  9.3 $^{+2.4}_{-1.9}$  & 0.629 $\pm 0.017$ & $10\pm72$ & Control & 1,53\\ 
XO-3b &  0.2833 $\pm 0.0034$  &  346.8 $^{+1.6}_{-1.5}$  &  0.2756 $\pm 0.0027$  &  -0.0649 $^{+0.0081}_{-0.008}$  &  1480 $\pm 11$  &  -0.019 $^{+0.025}_{-0.024}$  &  43.5 $^{+8.3}_{-7.0}$  & 12.15 $\pm 0.48$ & $37.3\pm3.0$ & Misaligned & 54,55\\ 
XO-4b &  0.002 $^{+0.012}_{-0.002}$  &  240 $^{+39}_{-160}$  &  0.00016 $^{+0.00062}_{-0.00051}$  &  -0.0001 $^{+0.0039}_{-0.0088}$  &  163.7 $\pm 4.7$  &  0.01 $\pm 0.01$  &  7.3 $^{+2.4}_{-1.9}$  & 1.559 $^{+0.052}_{-0.048}$ & $-46.7\pm8.1$ & Misaligned & 56\\ 
XO-5b &  0.013 $^{+0.014}_{-0.009}$  &  184 $\pm 92$  &  -0.0031 $^{+0.0073}_{-0.013}$  &  -0.0001 $^{+0.012}_{-0.013}$  &  144.3 $^{+2.9}_{-3.0}$  &  0.0041 $\pm 0.0029$  &  10.7 $^{+2.3}_{-1.8}$  & 1.051 $\pm 0.032$ & & Control & 57\\ 
\enddata
\tablenotetext{a}{Systems with accelerations that differ from zero by more than $3\sigma$ are marked in bold.}
\tablenotetext{b}{The misaligned sample consists of planets with either eccentric or misaligned orbits, the control sample contains planets that appear to have circular and/or well-aligned orbits.}
\tablenotetext{c}{Because the accelerations in the WASP-8 and WASP-34 systems have some curvature, we fit then with a two-planet solution where the linear trend slope term is fixed to zero (see Tables \ref{wasp8_table} and \ref{wasp34_table} for the full solution).}
\tablenotetext{d}{Our values differ from those of previous fits, which did not include the measured secondary eclipse times.}
\tablenotetext{e}{REFERENCES FOR ORBITAL INCLINATIONS AND SPIN-ORBIT ANGLES  -
(1) \citet{torres08}; (2) \citet{carter09}; (3) \citet{albrecht12b}; (4) \citet{pal10}; (5) \citet{winn11}; (6) \citet{winn09}; (7) \citet{simpson11}; (8) \citet{moutou11}; (9) \citet{west09b}; (10) \citet{bakos10}; (11) \citet{winn10c}; (12) \citet{hartman09}; (13) \citet{winn10a}; (14) \citet{torres10}; (15) \citet{winn11}; (16) \citet{kovacs10}; (17) \citet{buchhave10}; (18) \citet{howard12}; (19) \citet{fulton13}; (20) \citet{hartman11a}; (21) \citet{bakos11}; (22) \citet{kipping10}; (23) \citet{hartman11b}; (24) \citet{buchhave11}; (25) \citet{johnson11}; (26) \citet{kipping11}); (27) \citet{hartman11c}; (28) \citet{bakos12}; (29) \citet{winn08}; (30) \citet{sozzetti09};  (31) \citet{mandushev07}; (32) \citet{narita10}; (33) \citet{simpson11}; (34) \citet{albrecht11}; (35) \citet{triaud10}; (36) \citet{gibson08}; (37) \citet{tripathi10}; (38) \citet{sanchis11}; (39) \citet{hellier08}; (40) \citet{queloz10}; (41) \citet{johnson09b}; (42) \citet{maciejewski11a}; (43) \citet{joshi09}; (44) \citet{johnson09}; (45) \citet{lister09}; (46) \citet{anderson10}; (47) \citet{hellier11}; (48) \citet{tregloan12}; (49) \citet{maxted10}; (50) \citet{anderson11}; (51) \citet{smalley11}; (52) \citet{brown12b}; (53) \citet{narita11}; (54) \citet{johnskrull08}; (55) \citet{hirano11}; (56) \citet{narita10}; (57) \citet{maciejewski11b}
}
\end{deluxetable}
\clearpage
\end{landscape}

We computed 2$\times N$ DE-MCMC chains (where $N$ is the number of free parameters in the RV model), continuously checking for convergence following the prescription of \citet{eastman13}.  We considered the chains well-mixed and halted the DE-MCMC run when the number of independent draws \citep[$T_{z}$, as defined in][]{ford06} was greater than 1000 and the Gelman-Rubin statistic \citep{gelman03,ford06,holman06} was within 1\% of unity for all parameters.  In order to speed convergence (however see section \ref{sec:correlations}), ensure that all parameter space was adequately explored, and minimize biases in parameters that physically must be finite and positive, we step in the widely used modifications and/or combinations of orbital parameters: log(P), \ecosw , \esinw , and log(K).

We assigned Gaussian priors to P, T$_{\rm mid}$ and secondary eclipse times where available as listed in Tables \ref{rv_priors1} and \ref{rv_priors2}, and we assigned uniform priors to all other parameters. The reference epoch (abscissa) for $\dot{\gamma}$ was chosen as the mid-time of the RV time-series in order to minimize the covariance between $\gamma$ and $\dot{\gamma}$.  In all cases we assumed that transit timing variations caused by other known or unknown companions were negligible.  The median parameter values and associated 68\% confidence intervals from the DE-MCMC analysis for all systems are presented in Tables \ref{rv_params1} and \ref{rv_params2}.

\subsection{Parameter correlations}\label{sec:correlations}

While vetting the fits for all planets we noticed that in some cases the two dimensional marginalized distributions when plotted in \esinw\ vs. \ecosw\ took on non-Gaussian shapes in systems for which we had many secondary eclipse times to constrain the orbital phase of the secondary eclipse (and thus $e\cos{\omega_{\star}}$). Systems for which we had good secondary eclipse priors but the eccentricity was low took on a star-shaped appearance while a strong correlation between \esinw\ and \ecosw\ emerged in systems with significant eccentricity.  This correlation was much less pronounced when we plotted $e\cos{\omega_{\star}}$ versus $e\sin{\omega_{\star}}$ (Figure \ref{fig:distros}). In order to check that our DE-MCMC algorithm was behaving as expected we created hypothetical distributions of $e$ and $\omega$ for two cases. 

\begin{figure}[ht]
\epsscale{1.2}
\plotone{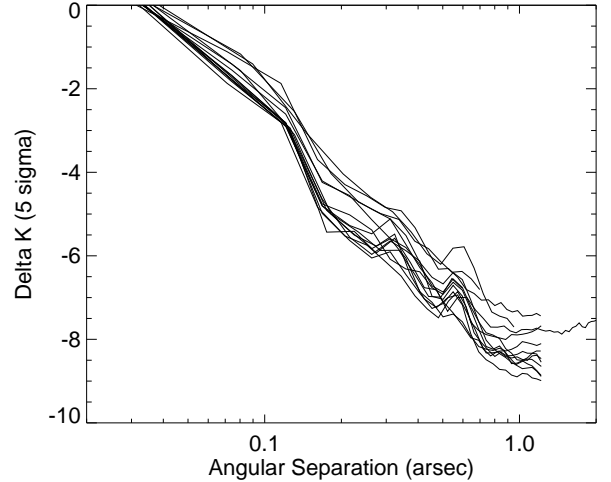}
\centering
\caption{$K_p$ and $K_s$ contrast curves for targets with radial velocity trends.  HAT-P-20 is one of our closest targets, and we therefore used non-overlapping regions from our nodded images to extend our effective field of view to larger separations.  Additional details on the images used to calculate these contrast curves can be found in Table \ref{ao_obs}.}
\label{contrast_curve_plot}
\end{figure}

First, we created a hypothetical distribution of $e$ as the absolute value of a Gaussian centered around zero and a uniform distribution of $\omega$ between 0 and $2\pi$ and to simulate the posterior distributions for a planet with no significant eccentricity. We then extracted the points within the distributions for which the orbital phase of the secondary eclipse calculated from $e$ and $\omega$ was very close the median value of all secondary eclipse times calculated from the $e$ and $\omega$ distributions (near phase=0.5 in this case). When the entire distribution is plotted in \esinw\ vs. \ecosw\ we see a smooth distribution with circular contours, but when the points extracted based on the secondary eclipse times are plotted we see the star shape that closely resembles the distributions we obtain in our fits to the data.

Second, we created a hypothetical distribution of $e$ as a normal distribution centered around a value of 0.16 with a width of 0.02 and a normal distribution of $\omega$ centered around 328 degrees with a width of 10 degrees. These distributions are meant to mimic a planet with significant eccentricity. When points are extracted based on the secondary eclipse times in the same way as the first case we see that the points fall along a locus that matches the distributions obtained in our fits.

This test shows that our DE-MCMC algorithm is working as expected, but that the choice of parameterization to use \esinw and \ecosw~may not be optimal for systems with secondary eclipse measurements. However, since our DE-MCMC code continuously checks for convergence we know that the chains are converged and well-mixed. The correlated parameters will slow the convergence, but we decided that the factor of five increase in runtime was acceptable and did not change parameterization.

\subsection{Contrast Curves from AO Imaging}\label{ao_contrast}

We use our K band NIRC2 imaging data to place upper limits on the allowed masses and orbital semi-major axes of the perturbers responsible for the measured radial velocity accelerations.  Contrast curves are generated for each target as follows.  First, we calculate the full width at half max (FWHM) of the central star's point spread function in the interpolated and combined image.  The maximum radius for our contrast curves is defined as the largest radial separation for which data is available at all position angles (i.e., we do not count the corners of the array).  We then create a box with dimensions equal to the FWHM and step it across the array, calculating the total flux from the pixels within the box at a given position.  We exclude boxes containing masked pixels\footnote{We mask out regions containing detectable flux from nearby candidate stellar companions.  Objects with nearby companions include:  HAT-P-7 \citep{narita12}, HAT-P-10 (Ngo et al. in prep), HAT-P-32 \citep{adams13}, and WASP-8 \citep{queloz10}.} and boxes whose radial distance from the star is greater than our maximum radius limit. The $5\sigma$ contrast limit is calculated as a function of radial separation from the star by taking the standard deviation of the total flux values for boxes within a given annulus with width equal to the full width at half max of the stellar pdf (i.e., one box width) and multiplying by five.  We convert our absolute flux limits to relative delta magnitude units by taking the maximum flux value in the interpolated stellar point spread function as an estimate of the flux of the central star and calculating the corresponding relative magnitude limits for each radial distance.  We show the resulting contrast curves for all of our targets in Fig.~\ref{contrast_curve_plot}.

With the exception of GJ 436 and HAT-P-2, none of our target stars have directly measured parallax estimates.  In most cases the discovery paper provides an estimate of the stellar properties (mass, radius, and age) from fitting stellar evolution models using constraints on the surface gravity, effective temperature, and metallicity from high-resolution optical spectroscopy and (in some cases) constraints on the stellar density from fits to the transit light curve.  The distance can then be estimated using the known stellar properties and the measured apparent magnitudes in $V$, $J$, $H$, and $K$ bands.  We take these estimated distances and use them to convert the units of our contrast curves from separations in arc seconds to projected physical distances in AU

We convert our contrast curves from delta magnitudes in either $K_s$ or $K_p$ bands to mass limits for stellar companions using the latest version of the PHOENIX stellar atmosphere models \citep{husser13}.  We assume solar metallicities for both the primary and secondary, and interpolate in the available grid of models to produce a model that exactly matches the effective temperatures and surface gravities of each star.  We utilize the published temperatures and surface gravities for our primary stars, taking the best available constraints in each case.  We then systematically step through the table of radius and effective temperature as a function of secondary mass for a low-mass main-sequence companion from \citet{baraffe98} and create matching PHOENIX models for a corresponding secondary stellar companion with those properties.  The corresponding contrast ratio between the primary and secondary as a function of mass is calculated by integrating over the appropriate bandpass (either $K_p$ or $K_s$).  Finally, we convert our contrast curves from units of delta magnitude to secondary mass using the mass versus delta magnitude relations derived for that system.  

\begin{figure}[ht]
\epsscale{1.2}
\plotone{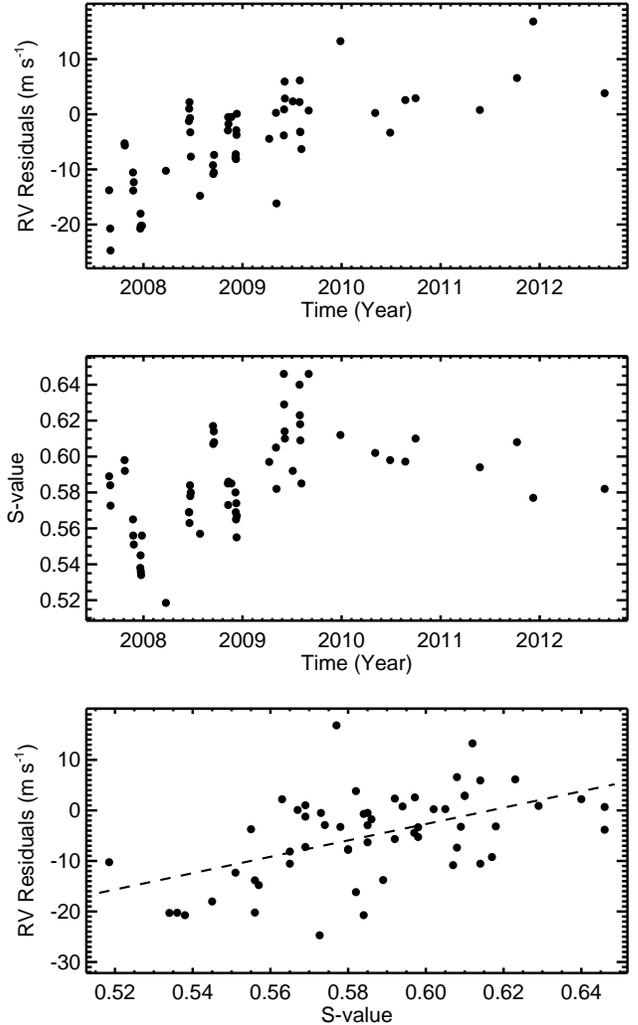}
\centering
\caption{Comparison of radial velocity trend and stellar activity index \sval~for the HAT-P-11 system.  Top panel:  radial velocity residuals after removing the best-fit orbital solution for the inner  transiting planet are shown as black filled circles.  Middle panel: activity index \sval~corresponding to each of the radial velocity measurements in the top panel.  Bottom panel: radial velocity residuals plotted as a function of \sval, with a linear fit shown as a black dashed line for comparison.}
\label{hat11_sval}
\end{figure}

\begin{figure*}[ht]
\epsscale{1.2}
\plotone{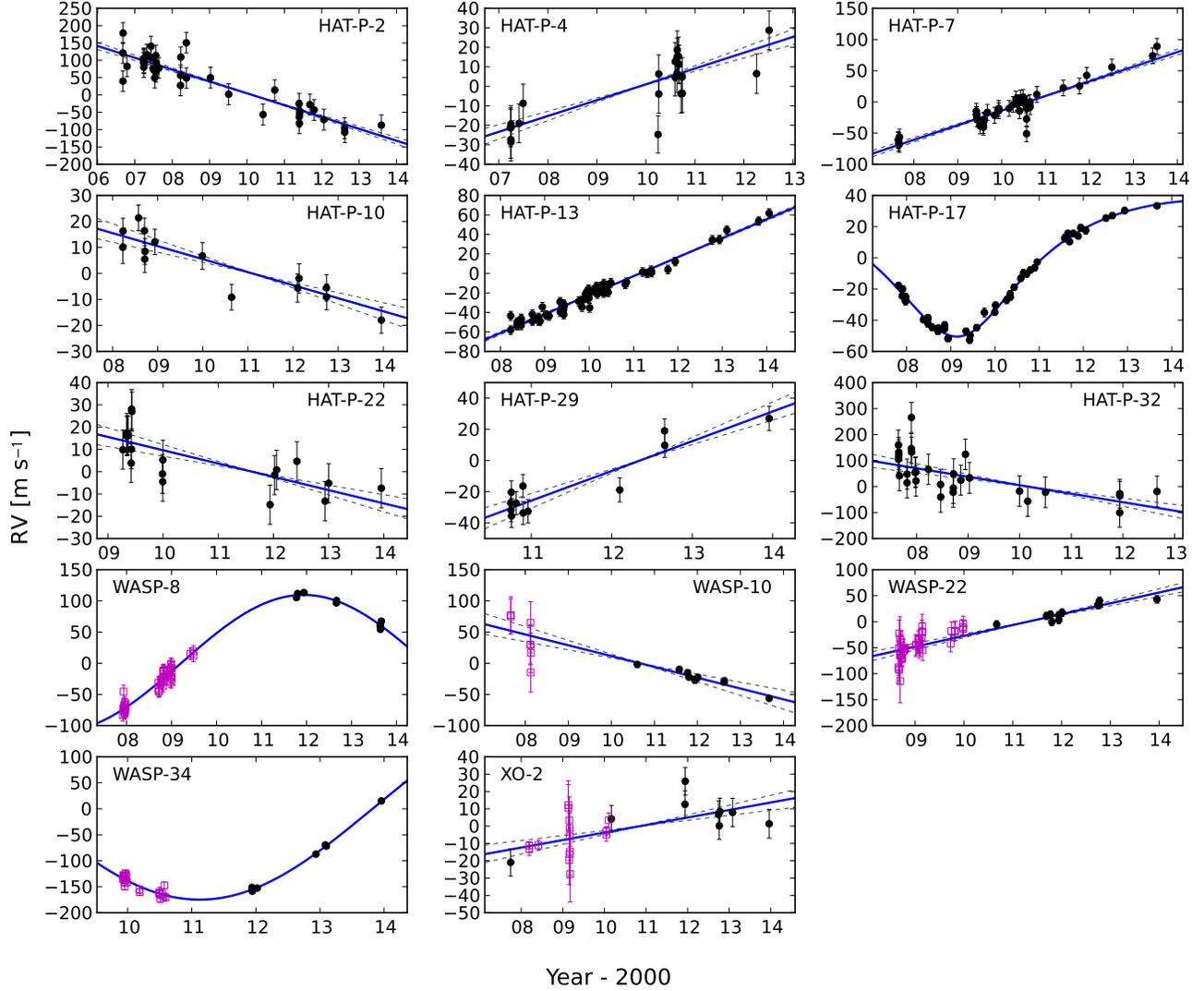}
\centering
\caption{Radial velocity data and best-fit accelerations for fifteen systems with known transiting gas giant planets and evidence for an outer companion whose orbit is not fully resolved.  The radial velocity signals from the transiting planets have been subtracted from the data shown in this plot.  The middle planet in the HAT-P-13 system, which has a complete orbit, is shown separately in Fig.~\ref{hat13_plot} and is also subtracted from this plot.  HIRES measurements are shown as black filled circles and measurements from other telescopes are shown as open purple squares.  Best-fit linear radial velocity accelerations are shown as a blue solid line, with $1\sigma$ errors as dashed grey lines.  The accelerations in the HAT-P-17, WASP-8 and WASP-34 systems all exhibit some curvature, and in these cases we over plot the best-fit solution for a companion with a circular orbit in blue (see Tables \ref{wasp8_table}, \ref{wasp34_table}, and \citet{fulton13} for more details on these systems).}
\label{rv_accelerations}
\end{figure*}

Our approach differs from the standard approach for AO imaging searches for stellar companions \citep[e.g.,][]{bechter13}, which typically utilize relative K magnitude estimates from 2MASS and parallax measurements to calculate an absolute K magnitude  for the primary and then interpolate in a grid of absolute K magnitudes as a function of secondary mass calculated from standard stellar evolution models at a given age \citep[e.g.,][]{girardi02}.  Our method offers two advantages over this approach:  first, we do not need a distance estimate to calculate the contrast ratio between the primary and secondary, and second, we can calculate contrast ratios in arbitrary bandpasses as needed.  We validate our method by converting the K band contrast curves for HAT-P-8 and WASP-12b from \citet{bechter13} using our new method, and find results that are consistent to within 0.02 solar masses in both cases.

\section{Discussion}\label{discussion}

\subsection{Trend Detections}\label{trend_discussion}

We find linear or curved trends in the measured radial velocities with slopes at least $3\sigma$ away from zero for fifteen systems listed in Tables \ref{rv_params1} and \ref{rv_params2}.  We next checked these systems to determine if any of the radial velocity trends were well-correlated with the stellar \caii~emission index \sval.  We find that one system, HAT-P-11, does exhibit a correlation with the measured \sval~and therefore conclude that this signal is likely the result of stellar activity rather than a real companion (see Fig. \ref{hat11_sval}).  This is not surprising, as this is one of the most active stars in our sample with a \logr~value of $-4.57$ \citep{knutson10}.  Of the remaining fourteen systems with evidence for an outer companion, eight have previously been reported in the published literature including: HAT-P-2 \citep{lewis13}, HAT-P-4 \citep{winn11}, HAT-P-7 \citep{winn09}, HAT-P-13 \citep{bakos09b,winn10a}, HAT-P-17 \citep{howard12,fulton13}, WASP-8 \citep{queloz10}, WASP-22 \citep{maxted10,anderson11}, and WASP-34 \citep{smalley11}.  We present a composite plot showing all of the detected radial velocity accelerations in Fig. \ref{rv_accelerations}. 

We also report new trend detections for six systems including: HAT-P-10, HAT-P-22, HAT-P-29, HAT-P-32, WASP-10, and XO-2.  Finally, we do not find statistically significant accelerations in the following systems with previously reported trend detections: GJ 436 \citep{maness07}, HAT-P-31 \citep{kipping11}, and HAT-P-34 \citep{bakos12}.  We discuss the differences between our results and those of previous studies for individual systems in \S\ref{prev_studies} below.

\subsubsection{Comparison to Previously Studies}\label{prev_studies}

Our values for the trend in the HAT-P-2 system are consistent with but less precise than those reported in \citet{lewis13}, although we are fitting the same radial velocity data set in both cases.  This is due to our treatment of the high-cadence data obtained as part of the Rossiter measurement for this system.  While Lewis et al. chose to give each out-of-transit Rossiter point equal weight in the fits, we took the error-weighted mean of the data from this observation and incorporated that averaged point in our fit.  Because we add a constant jitter term to all points, this effectively down-weights the contribution of the Rossiter data to our determination of the stellar slope.  Although this is a more conservative strategy that results in larger uncertainties on the best-fit trend slope, it effectively ensures that our fits are not biased by short-term trends caused by stellar activity and other sources of variability.  

\begin{deluxetable}{lrr}
\tablecaption{Fit Parameters for HAT-P-13 System \label{hat13_table}}
\tablewidth{0pt}
\tabletypesize{\tiny}
\tablehead{
\colhead{Parameter} & \colhead{Value} & \colhead{Units}
}
 \startdata
    \sidehead{RV Step Parameters}log($P_{b}$) & 0.46482299 $^{+3.2e-07}_{-3.3e-07}$ & log(days)\\
$T_{c,b}$ & 2455176.53877 $\pm 0.00027$ & \bjdtdb\\
$\sqrt{e_{b}}\cos{\omega_{b}}$ & -0.096 $^{+0.027}_{-0.021}$ & \\
$\sqrt{e_{b}}\sin{\omega_{b}}$ & -0.031 $^{+0.069}_{-0.059}$ & \\
log($K_{b}$) & 2.0248 $\pm 0.0032$ & \ms\\
log($P_{c}$) & 2.64916 $^{+0.00011}_{-0.0001}$ & log(days)\\
$T_{c,c}$ & 2455311.82 $\pm 0.19$ & \bjdtdb\\
$\sqrt{e_{c}}\cos{\omega_{c}}$ & -0.8068 $\pm 0.0013$ & \\
$\sqrt{e_{c}}\sin{\omega_{c}}$ & 0.0649 $\pm 0.0031$ & \\
log($K_{c}$) & 2.631 $\pm 0.0021$ & \ms\\
$\gamma$ & -23.04 $^{+0.84}_{-0.86}$ & \ms\\
$\dot{\gamma}$ & 0.0528 $^{+0.0013}_{-0.0014}$ & \ms day$^{-1}$\\
jitter & 4.53 $^{+0.54}_{-0.46}$ & \ms\\
\sidehead{RV Model Parameters}$P_{b}$ & 2.9162381 $^{+2.1e-06}_{-2.2e-06}$ & days\\
$T_{c,b}$ & 2455176.53877 $\pm 0.00027$ & \bjdtdb\\
$e_{b}$ & 0.0133 $^{+0.0047}_{-0.0044}$ & \\
$\omega_{b}$ & 197 $^{+32}_{-37}$ & degrees\\
$K_{b}$ & 105.87 $\pm 0.78$ & \ms\\
$P_{c}$ & 445.82 $\pm 0.11$ & days\\
$T_{c,c}$ & 2455311.82 $\pm 0.19$ & \bjdtdb\\
$e_{c}$ & 0.6551 $\pm 0.0021$ & \\
$\omega_{c}$ & 175.40 $\pm 0.22$ & degrees\\
$K_{c}$ & 427.6 $\pm 2.1$ & \ms\\
$\gamma$ & -23.04 $^{+0.84}_{-0.86}$ & \ms\\
$\dot{\gamma}$ & 0.0528 $^{+0.0013}_{-0.0014}$ & \ms day$^{-1}$\\
jitter & 4.53 $^{+0.54}_{-0.46}$ & \ms\\
\sidehead{RV Derived Parameters}$e\cos{\omega}$ & -0.0107 $^{+0.0039}_{-0.0041}$ & \\
$e\sin{\omega}$ & -0.0032 $^{+0.0073}_{-0.0077}$ & \\
$M_c\sin{i_c}$ & 14.61 $^{+0.46}_{-0.48}$ & \mj \\
$a_c$ & 1.258 $\pm 0.020$ & AU \\
 \enddata
\end{deluxetable}

\begin{figure}[ht]
\epsscale{1.2}
\plotone{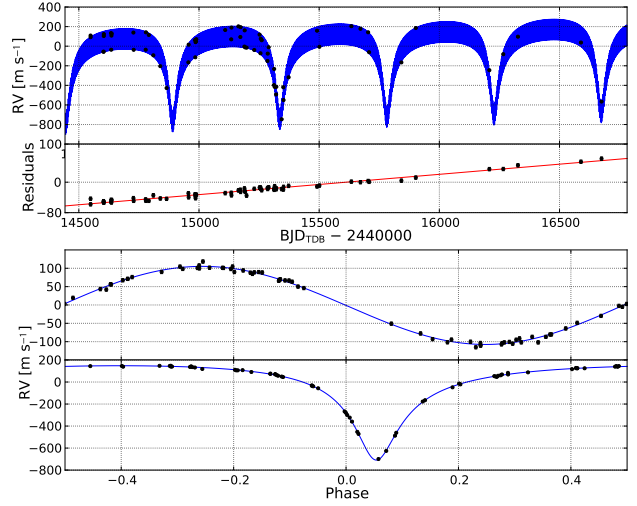}
\centering
\caption{Radial velocity measurements and best-fit curves for the HAT-P-13 system.  Top panel:  full radial velocity fit including two planets and a linear trend.  Top middle panel: residuals after accelerations from the two inner planets have been removed.  Lower middle panel:  phased radial velocity curve for the inner (b) transiting planet.  Bottom panel: phased radial velocity curve for the middle (c) planet.}
\label{hat13_plot}
\end{figure}

Our treatment of the Rossiter data sets affects our slope estimates for several other systems in addition to HAT-P-2.  For HAT-P-4 we find a trend slope consistent with the value of $0.0246\pm0.0026$ \ms~day$^{-1}$ reported in \citet{winn11}, but our slope has errors that are approximately twice as large as those reported by Winn et al.  Although we extend the baseline of the Winn et al. measurement from approximately 1300 to 1900 days, we also allow the eccentricity of the transiting planet to vary as a free parameter in our fits.  Winn et al. assume a circular orbit for the inner planet, which they find reduces the uncertainty on their estimate of the trend slope.  For HAT-P-7 we extend the approximately 600 day baseline from \citet{winn09} to 2100 days, and find a slope that is approximately $2\sigma$ larger with comparable uncertainties to those reported by Winn et al.  We find a similar situation for WASP-22 and WASP-34, where our best-fit slopes are consistent with the previously reported values but with errors that are factors of $1.5-2$ larger.  In all cases the planets in question had Rossiter observations spanning multiple hours that were included in the published fits to determine the trend slopes.  In our new fits we exclude the in-transit measurements and bin the out-of-transit measurements from each Rossiter observation into a single point in order to minimize the effect of short-term stellar jitter on our results.

Next we examine systems with previously reported trends that did not appear at a statistically significant level in our study.  The trend for GJ 436 from \citet{maness07} was only marginally significant ($3.4\sigma$), and this study assumed a stellar jitter value of 1.9 m s$^{-1}$ based on results from a sample of similar M stars.  In our study we extend the previous radial velocity baseline by eight years and fit for the jitter as a free parameter.  We find that the data prefer a value of $3.8\pm0.3$ m s$^{-1}$; if \citet{maness07} had used this jitter value their trend detection would have been below the threshold for statistical significance.   The non-detection of the trend in the HAT-P-31 system is puzzling, as this signal was detected with high statistical confidence in the original data set \citep{kipping11}.  Our new observations show no evidence of the curved trend visible in the original plots; in hindsight we suspect this curved fit was driven by a combination of two particularly low HIRES points and the use of data from three telescopes with limited sampling in individual data sets.  It is worth noting that Kipping et al. found a stellar jitter level of less than 2 m s$^{-1}$ in their fits, whereas we prefer a value of $6.6\pm1.5$ m s$^{-1}$.  We find no evidence for a correlation between the radial velocity residuals and the stellar activity index \sval; the primary star in this system has an effective temperature of approximately 6100 K, \vsini~less than 0.5 km s$^{-1}$, and \sval~equal to -5.3, making it unlikely that activity-induced jitter could have led to a spurious signal.  However, this does not preclude other sources of jitter.  The trend in the HAT-P-34 system \citep{bakos12} had a significance of $1.9\sigma$ and was based on just three months of radial velocity data; our observations span two years, and allow us to exclude the marginal slope reported in the original study.

\begin{deluxetable}{lrr}
\tablecaption{Fit Parameters for WASP-8 System \label{wasp8_table}}
\tablewidth{0pt}
\tabletypesize{\tiny}
\tablehead{
\colhead{Parameter} & \colhead{Value} & \colhead{Units}
}
 \startdata
    \sidehead{RV Step Parameters}log($P_{b}$) & 0.91162223 $^{+7.6e-07}_{-7.8e-07}$ & log(days)\\
$T_{c,b}$ & 2454679.33392 $\pm 0.00047$ & \bjdtdb\\
$\sqrt{e_{b}}\cos{\omega_{b}}$ & 0.04055 $\pm 0.00064$ & \\
$\sqrt{e_{b}}\sin{\omega_{b}}$ & -0.5502 $^{+0.0037}_{-0.0036}$ & \\
log($K_{b}$) & 2.3446 $\pm 0.0023$ & \ms\\
log($P_{c}$) & 3.636 $^{+0.068}_{-0.039}$ & log(days)\\
$T_{c,c}$ & 2452613 $^{+330}_{-610}$ & \bjdtdb\\
$\sqrt{e_{c}}\cos{\omega_{c}}$ & $\equiv$ 0.0 $\pm 0.0$ & \\
$\sqrt{e_{c}}\sin{\omega_{c}}$ & $\equiv$ 0.0 $\pm 0.0$ & \\
log($K_{c}$) & 2.061 $^{+0.062}_{-0.036}$ & \ms\\
$\gamma_{1}$ & -57.9 $^{+9.6}_{-17.0}$ & \ms\\
$\gamma_{2}$ & -20 $^{+24}_{-37}$ & \ms\\
$\gamma_{3}$ & 0 $^{+24}_{-38}$ & \ms\\
$\dot{\gamma}$ & $\equiv$ 0.0 $\pm 0.0$ & \ms day$^{-1}$\\
jitter & 2.91 $^{+0.4}_{-0.34}$ & \ms\\
\sidehead{RV Model Parameters}$P_{b}$ & 8.158724 $^{+1.4e-05}_{-1.5e-05}$ & days\\
$T_{c,b}$ & 2454679.33392 $\pm 0.00047$ & \bjdtdb\\
$e_{b}$ & 0.3044 $^{+0.0039}_{-0.004}$ & \\
$\omega_{b}$ & 274.215 $^{+0.084}_{-0.082}$ & degrees\\
$K_{b}$ & 221.1 $\pm 1.2$ & \ms\\
$P_{c}$ & 4323 $^{+740}_{-380}$ & days\\
$T_{c,c}$ & 2452613 $^{+330}_{-610}$ & \bjdtdb\\
$e_{c}$ & $\equiv$ 0.0 $\pm 0.0$ & \\
$\omega_{c}$ & $\equiv$ 90.0 $\pm 0.0$ & degrees\\
$K_{c}$ & 115.0 $^{+18.0}_{-9.2}$ & \ms\\
$\gamma_{1}$ & -57.9 $^{+9.6}_{-17.0}$ & \ms\\
$\gamma_{2}$ & -20 $^{+24}_{-37}$ & \ms\\
$\gamma_{3}$ & 0 $^{+24}_{-38}$ & \ms\\
$\dot{\gamma}$ & $\equiv$ 0.0 $\pm 0.0$ & \ms day$^{-1}$\\
jitter & 2.91 $^{+0.4}_{-0.34}$ & \ms\\
\sidehead{RV Derived Parameters}$e\cos{\omega}$ & 0.02237 $^{+0.00032}_{-0.00031}$ & \\
$e\sin{\omega}$ & -0.304 $\pm 0.004$ & \\
$M_c\sin{i_c}$ & 9.45 $^{+2.26}_{-1.04}$ & \mj \\
$a_c$ & 5.28 $^{+0.63}_{-0.34}$ & AU \\
 \enddata
\end{deluxetable}

\begin{figure}[ht]
\epsscale{1.2}
\plotone{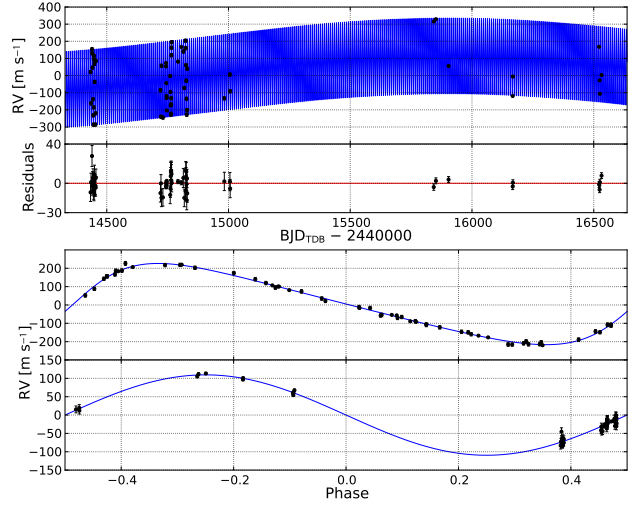}
\centering
\caption{Radial velocity measurements and best-fit curves for the WASP-8 system.  Top panel:  full radial velocity fit including two planets and a linear trend.  Top middle panel: residuals after accelerations from the two planets have been removed.  Lower middle panel:  phased radial velocity curve for the inner (b) transiting planet.  Bottom panel: phased radial velocity curve for the outer (c) planet.}
\label{wasp8_plot}
\end{figure}

\subsubsection{Fits to Systems With Curved Radial Velocity Trends or Multiple Planets}

We find three systems with evidence for curvature in the radial velocity acceleration, including: HAT-P-17, WASP-8, and WASP-34.  For HAT-P-13, the second planet has a fully resolved orbit and there is an additional linear acceleration present.  Our fits for HAT-P-17 follow the methods described in \citet{fulton13} and we obtain values that are consistent with that study; this is not surprising, as we have added just two new radial velocity measurements in our fits.  We discuss our results for the other three systems individually below.

\citet{queloz10} reported a linear trend with a slope of $58.1\pm1.3$ m s$^{-1}$ yr$^{-1}$ for the WASP-8 system.  We find that this trend has turned over in our new observations, allowing us to fit for the orbital properties of the outer companion rather than assuming a linear trend.  We show our results from these new fits in Table \ref{wasp8_table} and Fig.~\ref{wasp8_plot}.  We find that when we assume a circular orbit for the outer companion we obtain a reasonably well-constrained orbital solution with a period of $4339^{+850}_{-390}$ days, a radial velocity semi-amplitude $K$ of $115.7^{+21.0}_{-9.4}$, and $M \sin i$ equal to 9.5 $^{+2.7}_{-1.1}$ $\mj$.  When we allow the eccentricity to vary freely in the fits our chains do not converge on a well-defined solution, and both the period and radial velocity semi-amplitude span a larger range in values ($3971^{+1100}_{-900}$ days and $110^{+24}_{-27}$ m s$^{-1}$, respectively).  We present the results for the better-constrained circular fit in Table \ref{wasp8_table} and consider the non-circular case in more detail in \S\ref{rv_companions}; we note that the data are equally consistent with both circular and eccentric fits.

\begin{figure}[ht]
\epsscale{1.2}
\plotone{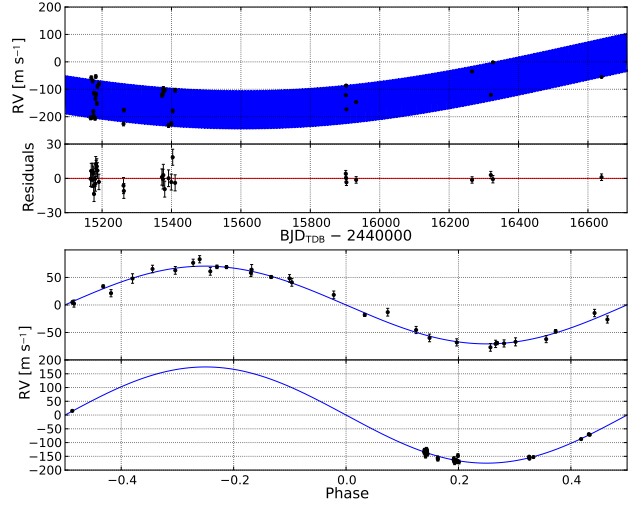}
\centering
\caption{Radial velocity measurements and best-fit curves for the WASP-34 system.  Top panel:  full radial velocity fit including two planets and a linear trend.  Top middle panel: residuals after accelerations from the two planets have been removed.  Lower middle panel:  phased radial velocity curve for the inner (b) transiting planet.  Bottom panel: phased radial velocity curve for the outer (c) planet.}
\label{wasp34_plot}
\end{figure}

Our new measurements also indicate that the linear trend reported in \citet{smalley11} for WASP-34 has turned over, allowing us to  place weak constrains on the orbital period and mass of the companion for the case of a circular orbit (see Table \ref{wasp34_table} and Fig. \ref{wasp34_plot}).  We find that in these fits the companion has an orbital period of $4133^{+2500}_{-1400}$ days, a radial velocity semi-amplitude $K$ of $196^{+280}_{-98}$, and $M \sin i$ equal to 15.5 $^{+28.0}_{-8.8}$ $\mj$.  We evaluate the constraints for the non-circular case separately in \S\ref{rv_companions}.  

HAT-P-13 presents a particularly interesting case, as our fits indicate evidence for two outer companions in the system.  Previous studies \citep{bakos09b,winn10a} reported the presence of one companion with a fully resolved orbit (``c") and an additional radial velocity trend (``d").  We provide an updated estimate for the properties of companion ``c", which has a period of $445.87\pm0.12$ days and an orbital eccentricity of $0.6573\pm0.0034$.  We estimate a $M \sin i$ of 14.70 $^{+0.48}_{-0.47}$ $\mj$ for this companion, and list the full set of fit parameters in Table \ref{hat13_table}.  Dynamical studies of this system \citep{batygin09,becker13} predict that this planet will perturb on the orbit of the inner hot Jupiter (``b"), resulting in the alignment of the apses of the two planetary orbits.  In this scenario, the eccentricity of the inner planet can be used to constrain the inner planet's tidal Love number (a measure of its degree of central concentration).   Our new fits indicate that the arguments of periapse $omega$ for the orbits of the two inner planets are consistent but with large uncertainties on $omega_b$, which are primarily due to this planet's small orbital eccentricity \citep[also see ][]{winn10a}.  We show the phased radial velocity curves in Fig.~\ref{hat13_plot}, and provide additional constraints on the properties of companion ``d" in \S\ref{rv_companions}.  When we include a second planet in our fit to HAT-P-17 we obtain the same orbital parameters as those reported in \citep{fulton13}, with no evidence for any additional accelerations in this system.

\begin{deluxetable}{lrr}
\tablecaption{Fit Parameters for WASP-34 System \label{wasp34_table}}
\tablewidth{0pt}
\tabletypesize{\tiny}
\tablehead{
\colhead{Parameter} & \colhead{Value} & \colhead{Units}
}
 \startdata
    \sidehead{RV Step Parameters}log($P_{b}$) & 0.63525024 $\pm 4.5e-07$ & log(days)\\
$T_{c,b}$ & 2454647.55434 $^{+0.00063}_{-0.00064}$ & \bjdtdb\\
$\sqrt{e_{b}}\cos{\omega_{b}}$ & -0.002 $^{+0.062}_{-0.064}$ & \\
$\sqrt{e_{b}}\sin{\omega_{b}}$ & -0.02 $\pm 0.11$ & \\
log($K_{b}$) & 1.8517 $^{+0.0097}_{-0.01}$ & \ms\\
log($P_{c}$) & 3.612 $^{+0.073}_{-0.059}$ & log(days)\\
$T_{c,c}$ & 2454586 $^{+140}_{-190}$ & \bjdtdb\\
$\sqrt{e_{c}}\cos{\omega_{c}}$ & $\equiv$ 0.0 $\pm 0.0$ & \\
$\sqrt{e_{c}}\sin{\omega_{c}}$ & $\equiv$ 0.0 $\pm 0.0$ & \\
log($K_{c}$) & 2.28 $^{+0.12}_{-0.09}$ & \ms\\
$\gamma_{1}$ & 108 $^{+62}_{-37}$ & \ms\\
$\gamma_{2}$ & 141 $^{+62}_{-37}$ & \ms\\
$\dot{\gamma}$ & $\equiv$ 0.0 $\pm 0.0$ & \ms day$^{-1}$\\
jitter & 3.2 $^{+0.72}_{-0.6}$ & \ms\\
\sidehead{RV Model Parameters}$P_{b}$ & 4.3176779 $\pm 4.5e-06$ & days\\
$T_{c,b}$ & 2454647.55434 $^{+0.00063}_{-0.00064}$ & \bjdtdb\\
$e_{b}$ & 0.0109 $^{+0.015}_{-0.0078}$ & \\
$\omega_{b}$ & 215 $^{+77}_{-140}$ & degrees\\
$K_{b}$ & 71.1 $^{+1.6}_{-1.7}$ & \ms\\
$P_{c}$ & 4093 $^{+750}_{-520}$ & days\\
$T_{c,c}$ & 2454586 $^{+140}_{-190}$ & \bjdtdb\\
$e_{c}$ & $\equiv$ 0.0 $\pm 0.0$ & \\
$\omega_{c}$ & $\equiv$ 90.0 $\pm 0.0$ & degrees\\
$K_{c}$ & 189 $^{+60}_{-35}$ & \ms\\
$\gamma_{1}$ & 108 $^{+62}_{-37}$ & \ms\\
$\gamma_{2}$ & 141 $^{+62}_{-37}$ & \ms\\
$\dot{\gamma}$ & $\equiv$ 0.0 $\pm 0.0$ & \ms day$^{-1}$\\
jitter & 3.2 $^{+0.72}_{-0.6}$ & \ms\\
\sidehead{RV Derived Parameters}$e\cos{\omega}$ & -0.0001 $^{+0.0068}_{-0.0071}$ & \\
$e\sin{\omega}$ & -0.001 $^{+0.011}_{-0.017}$ & \\
$M_c\sin{i_c}$ & 14.96 $^{+6.29}_{-3.39}$ & \mj \\
$a_c$ & 5.05 $^{+0.65}_{-0.46}$ & AU \\
 \enddata
\end{deluxetable}

\subsection{AO Companions in Trend Systems}\label{ao_companions}

Four of our targets with detected radial velocity accelerations also host candidate AO-detected companions; in this section we consider whether such companions could explain the observed radial velocity trend.  For the cases where estimated spectral types are not available, we estimate the masses of the companions based on their brightness in K band relative to the primary using the same methods described in \S\ref{ao_contrast}.  We convert their projected separations on the sky to a minimum semi-major axis using the estimated distances of these systems, and then compare the estimated masses and minimum separations to the lower limit on the companion mass at that separation from the measured radial velocity trend.  We calculate this lower limit using the following expression \citep{torres99,liu02}:

\begin{align}\label{eq1}
M_{\rm comp} &=& 5.34\times 10^{-6} M_{\sun} \left(\frac{d}{\rm pc}\frac{\rho}{\rm arcsec}\right)^2 \nonumber \\
& & \times \left|\frac{\dot{v}}{\rm m\phantom{0}s^{-1}\phantom{0}yr^{-1}}\right|F(i,e,\omega,\phi)
\end{align}

where $d$ is the distance to the system, $\rho$ is the projected separation of the companion on the sky, $\dot{v}$ is the best-fit radial velocity trend, and $F(i,e,\omega,\phi)$ is an equation that depends on the unknown orbital parameters of the companion.  Liu et al. find that this equation has a minimum value of $\sqrt{27}/2$, which we use in our calculations here.

HAT-P-7 is known to have a common proper motion companion with a projected separation of $3.\arcsec9$ and spectral type of M5.5 \citep{narita12}, corresponding to a mass of approximately 0.2 M$_{\sun}$.  We convert this angular separation to a physical separation of 1240 AU using the estimated distance of $320^{+50}_{-40}$ pc from \citet{pal08}.  At this distance the minimum mass of the companion required to produce the measured radial velocity trend of $25.4\pm1.3$ m s$^{-1}$ yr$^{-1}$ is $540$ M$_\sun$.  We therefore conclude that the observed companion cannot be responsible for the trend in this system, in agreement with the conclusion of \citet{narita12}.

\begin{deluxetable}{l c c c c }
        \tabletypesize{\footnotesize}
\tablecaption{$1\sigma$ Constraints on Companion Properties\tablenotemark{a} \label{companion_table}}
\renewcommand{\arraystretch}{0.7}
\tablewidth{0pt}
\tablehead{
\colhead{Companion} & \colhead{M$_c \sin i$ (M$_{\rm Jup}$)} & \colhead{a (AU)} &  \colhead{Ref.}
} 
 \startdata
HAT-P-2c & $8-200$ & $4-31$ & 1,2 \\
HAT-P-4c & $1.5-310$ & $5-60$ & 2,3,4 \\ 
HAT-P-7c & $9-500$ &  $7-35$ & 5,6 \\  
HAT-P-10c\tablenotemark{b} & $>0.8$ & $>4.2$ & \\
HAT-P-13c\tablenotemark{c} & $14.23-15.18$ & $1.24-1.28$ & 2,7,8 \\
HAT-P-13d & $15-200$ & $12-37$ & 2,7,8 \\
HAT-P-17c\tablenotemark{d} & $2.8-3.7$ & $4.7-8.3$ & 2,9 \\ 
HAT-P-22c & $0.7-125$ & $3.0-28$ & 2,10 \\ 
HAT-P-29c & $1-200$ & $2-36$ & 2,11 \\ 
HAT-P-32c & $5-500$ & $3.5-21$ & 12\\ 
WASP-8c & $6.3-10.7$ &  $4.2-5.2$ & 2,13,14\\ 
WASP-10c & $4-90$ &  $5-30$ & 2,13,14\\ 
WASP-22c & $7-500$ & $6-40$ & 15 \\ 
WASP-34c & $28-98$ & $3.1-3.8$ & 16 \\ 
XO-2c & $0.6-70$ &  $3-23$ & 2,17 \\ 
\enddata
\tablenotetext{a}{We exclude HAT-P-11 from this list as the observed trend is likely the result of stellar activity.}
\tablenotetext{b}{HAT-P-10 has a directly imaged low-mass stellar companion that is consistent with the observed trend (see \S\ref{ao_companions}).}
\tablenotetext{c}{This planet is the only companion with a fully resolved orbit, and its parameters are taken directly from the fit presented in Table \ref{hat13_table}.}
\tablenotetext{d}{Also see \citet{fulton13}.}
\tablenotetext{e}{REFERENCES FOR STELLAR PROPERTIES  -
(1) \citet{pal10}; (2) \citet{torres12}; (3) \citet{kovacs07}; (4) \citet{winn11}; (5) \citet{pal08}; (6) \citet{vaneyken12}; (7) \citet{bakos09b}; (8) \citet{southworth12}; (9) \citet{howard12}; (10) \citet{bakos11}; (11) \citet{buchhave11}; (12) \citet{hartman11c}; (13) \citet{christian09}; (14) \citet{johnson09b}; (15) \citet{anderson11}; (16) \citet{smalley11}; (17) \citet{burke07}
}
\end{deluxetable}

\citet{adams13} reported the discovery of a candidate companion to HAT-P-32 with a projected separation of $2.\arcsec9$ and a delta magnitude of 3.4 in the $K_s$ band.  If we assume that this is a bound companion at the same distance as the primary, we find an estimated mass of 0.4 M$_\sun$.  We take our distance estimate of $285\pm5$ pc from \citet{hartman11c}, and calculate a corresponding physical separation of 830 AU for the companion.  We compare this value this to the minimum mass of $318$ M$_\sun$ required to explain the radial velocity trend slope of $-33\pm10$ m s$^{-1}$ yr$^{-1}$ at this separation and conclude that the companion cannot be responsible for the measured trend in this system.

\citet{queloz10} report a common proper motion companion to WASP-8 with a sky-projected separation of $4.\arcsec83\pm0.\arcsec01$ and a relative K magnitude of 2.1 from 2MASS photometry.  Assuming a distance of $87\pm7$ pc \citep{queloz10}, we find that this companion has a physical separation of 390 AU and an estimated mass of 0.5 M$_{\sun}$.  This is much less than the minimum mass of $125$ M$_{\sun}$ needed to explain the radial velocity trend of $58.1^{+1.2}_{-1.3}$ m s$^{-1}$ yr$^{-1}$ from Queloz et al.  We note that our new radial velocity data show a downward slope, indicating that the radial velocity trend reached a maximum in the past few years; this provides additional support for the hypothesis that the radial velocity trend is caused by a third, close-in body in the system.

Our preliminary $K$ band AO imaging also resulted in the detection of a previously unknown companion to HAT-P-10 with a sky-projected separation of $0.\arcsec34$ and a relative magnitude of $2.4$ in the $K_p$ band (Ngo et al. in prep).  Assuming a distance of $122\pm4$ pc from \citep{bakos09a}, we find that this corresponds to a sky-projected separation of 42 AU and a companion mass of $0.36$ M$_{\sun}$.  This is easily consistent with the minimum mass of $0.12$ M$_{\sun}$ needed to explain the measured radial velocity trend of $-5.1\pm1.4$ m s$^{-1}$ yr$^{-1}$; we therefore conclude that HAT-P-10 is the only system where an AO companion might explain the presence of the radial velocity trend, and we exclude this system from our subsequent analysis of the frequency of substellar companions in \S\ref{pop_stats}.

\subsection{Constraints on Companion Properties}\label{rv_companions}

We next simulate RV observations of each of our stars to determine what constraints we may place on the properties of the companions responsible for the observed radial velocity accelerations.  We refer to these plots as ``Wright diagrams" \citet{wright07}.  For each star, we develop a logarithmically spaced $50\times50$ grid of possible companion masses and semi-major axes spanning the range $0.2\, M_J < m \sin i < 500\,M_J$ and $1\,\textrm{AU} < a < 75\,\textrm{AU}$.  At each mass and semi-major axis, we inject a simulated planet with a fixed eccentricity and determine the orbital parameters which allow for the best fit to the RV observations. We calculate a $\chi^2$ value at each grid point for each eccentricity simulated, assuming our RV uncertainties (calculated as the quadrature sum of the reported errors and the best-fit jitter value) are random, uncorrelated, and Gaussian. We convert these likelihood values to a normalized probability, then marginalize over eccentricity. Here, we assume the long-period giant planet eccentricity distribution is well-replicated by the beta distribution \citep{kipping13}:

\begin{equation}
P_{\beta}(e;a,b)= \frac{\Gamma (a+b)}{\Gamma (a) \Gamma(b)} e^{a-1}(1-e)^{b-1}
\end{equation}

where $P_{\beta}$ is the probability of a given eccentricity, $\Gamma$ is a Gamma function, and $a$ and $b$ are constants that are fitted to the known population of long-period giant planets ($a=1.12$ and $b=3.09$ here).

\begin{figure*}[ht]
\epsscale{1.39}
\plotone{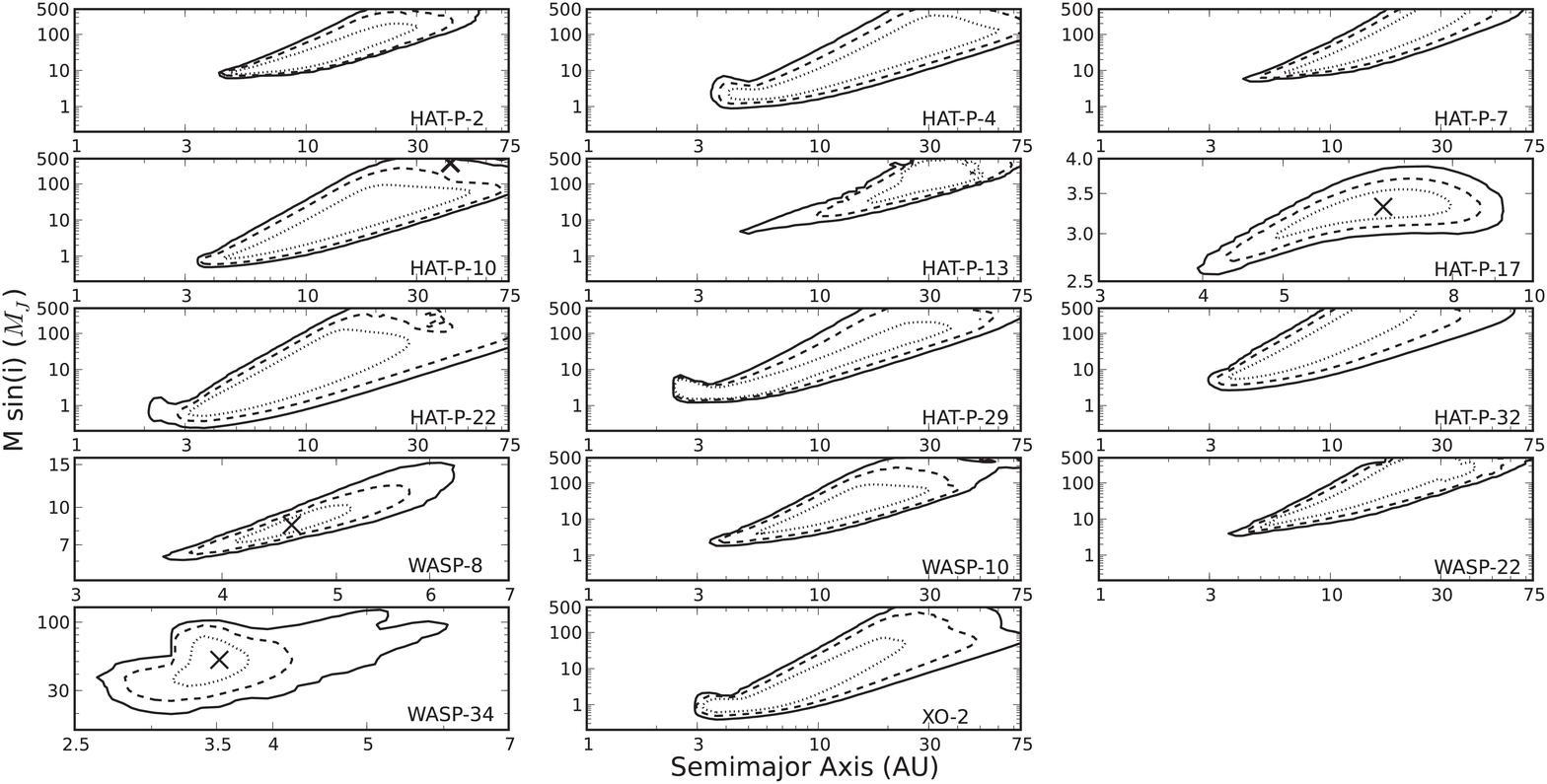}
\centering
\caption{``Wright diagrams"  showing the probability contours for the companions causing linear radial velocity accelerations.  We show the 1, 2, and 3$\sigma$ confidence intervals as dotted, dashed, and solid lines, respectively.  We exclude HAT-P-11, where the measured trend is likely due to stellar activity.  HAT-P-10 has a directly imaged companion with a mass and projected separation consistent with the measured radial velocity trend, which we indicate on this plot with an $X$ symbol.  We also include the constraints from our contrast curve for this system, which indicate that there are no other companions in this region of parameter space.  The companions to HAT-P-17, WASP-8, and WASP-34 display enough curvature in their measured radial velocities to provide strong constraints on their orbits, and we indicate the best-fit orbital solution for each companion with an $X$ in these plots.  The middle (``c") companion to HAT-P-13 has a fully resolved orbit with properties given in Table \ref{hat13_table}, and is not replotted here.  For all of these systems, we could, in theory, place additional constraints on very high mass, small separation companions from the lack of detectible lines in the visible-light HIRES spectra, but such companions are already disfavored by our current data.}
\label{companion_properties_plot}
\end{figure*}

If the trend is truly linear, we would not expect to break the degeneracy between the companion mass and semi-major axis, as the two are degenerate. For a given trend, the mass of a companion is proportional to the square of the companion's semi-major axis. In these cases, from the RV observations alone we can only place a lower limit on the mass and separation of a companion, as the orbit must be significantly longer than the RV baseline in order to produce a strictly linear trend. We place an outer limit on the companion's orbit using the contrast curves derived from our K-band AO imaging described in \S\ref{ao_contrast}, by injecting and ``imaging'' artificial companions in the same manner as described by \citet{montet13}.

A non-detection does not necessarily imply that the star does not host a giant companion; the companion may simply be too small or distant to induce a detectable RV acceleration. We can determine the likelihood of detecting a companion as a function of mass and semi-major axis. To accomplish this, we simulate planets over the grid described previously. For all stars, we inject 1000 planets at each mass and semi-major axis included in our grid. We randomly assign all other parameters following the distributions described above. We then ``observe'' each star by integrating the companion's orbit and calculating the magnitude of the RV signal at the times of our RV observations. Each velocity is then perturbed from the expected value by a normal variate with zero mean and standard deviation $\sigma$ equal to the RV uncertainty. If the best-fit to the RV acceleration is $3\sigma$ different from zero, we consider this companion to be detected.

The end result of this analysis is a map (the Wright diagram) showing either the range of companions in mass versus semi-major axis space that are consistent with the measured radial velocity acceleration, or a map showing the region of this space where companions are excluded by the current measurements (Fig.~\ref{companion_properties_plot}).  Although we also presented separate two-planet fits for the HAT-P-13, HAT-P-17 \citep[see][]{fulton13}, WASP-8, and WASP-34 systems, in the case of the latter two systems we assumed that the outer companion had a circular orbit in order to avoid degeneracies between the companion's eccentricity, mass, and semi-major axis in our fits.  Our new maps (Fig.~\ref{companion_properties_plot}) allow these partially resolved companions to have a non-zero orbital eccentricity, allowing for a broader range of orbital solutions.  We do not calculate a map for the middle (``c") companion in the HAT-P-13 system, as this planet has a fully resolved orbit with well-constrained properties listed in Table \ref{hat13_table}.  We list the constraints on each companion from this analysis in Table \ref{companion_table}.

\subsection{The Distribution of Wide Companions}\label{pop_stats}

We have determined the parameter space in mass and semi-major axis where a companion could reside for each of our fifteen systems exhibiting RV accelerations corresponding to companions whose orbits are not yet fully resolved.  We can combine this information to determine the most likely underlying giant companion distribution for systems hosting short period gas giant planets. 

We assume giant planets are distributed in planet mass and semi-major axis subject to the double power law
\begin{equation}
f(m,a) = C m^\alpha a^\beta d \ln m \, d \ln a.
\end{equation}

The likelihood of the data for a star with an RV trend detection is 
\begin{equation}
L_i  = \int d \ln m \int d \, \ln a \, f(m,a) \, p_i(m,a),
\end{equation}
where $p_i(m,a)$ is the probability of a planet at mass $m$ and orbital semi-major axis $a$, as calculated using the technique of the previous section. For HAT-P-13c we have a well constrained orbit, and we approximate $p_i(m,a) \approx \delta(m_i,a_i)$. In this case, the likelihood of the data is simply $L_i = f(m_i,a_i)$.

\begin{deluxetable*}{l|c|ccc}
        \tabletypesize{\footnotesize}
        \tablecolumns{5}
        \tablecaption{Maximally Likely Values for Distribution Parametrs.\tablenotemark{a}}
        \tablehead{ Parameter & Max $\mathcal{L}$ & $1 \sigma$ & $2\sigma$ & $3\sigma$}
        \startdata
  All Targets & & & & \\   
  $C$ & $7.6 \times 10^{-4}$ & [$1.2 \times 10^{-3}$,0.013] & [$2.3 \times 10^{-4}$, 0.031] & [$3.8 \times 10^{-5}$, 0.062]  \\ 
  $\alpha$ & 1.7 & [0.5, 1.6] & [0.1, 2.3] & [-0.2, 3.2] \\
  $\beta$ & 0.9 & [0.3, 1.0] & [-0.1, 1.4] & [-0.4, 1.8] \\
&&&& \\
  ``Misaligned''\tablenotemark{b} & & & & \\   
  $C$ & $1.2 \times 10^{-3}$ & [0.0016,0.018] & [$3.4 \times 10^{-4}$, 0.044] & [$7.0 \times 10^{-5}$, 0.09]  \\ 
  $\alpha$ & 1.8 & [0.6, 1.7] & [0.1, 2.4] & [-0.2, 3.1] \\
  $\beta$ &  0.6 & [-0.2, 0.6] & [-0.7, 1.1] & [-1.0, 1.5] \\
&&&& \\
  ``Control''\tablenotemark{b} & & & & \\   
  $C$ & $6.2 \times 10^{-4}$ & [0.0036,0.051] & [$5.1 \times 10^{-4}$ ,0.12] & [$7.3 \times 10^{-5}$,0.24]  \\ 
  $\alpha$ & 1.0 & [-0.5,0.6] & [-1.0,1.5] & [-1.4,2.5] \\
  $\beta$ & 1.3 & [0.1,1.3] & [-0.5,1.9] & [-1.1,2.6] \\
        \enddata
        \label{T:CAB}
        \tablenotetext{a}{$C$ is a normalization factor which depends on the total giant planet occurrence rate, $\alpha$ is the power law exponent for frequency as a function of mass, and $\beta$ is the power law exponent for frequency as a function of semi-major axis. Because some distributions are significantly skewed, the maximum likelihood value is not included inside the $1\sigma$ confidence interval for all parameters.}
        \tablenotetext{b}{See Table \ref{rv_params} for a list of the systems included in the ``Misaligned'' and ``Control'' samples, and \S\ref{pop_stats} for a description of how the two samples were devised.}
\end{deluxetable*}

For a non-detection, we are able to rule out high-mass, close-in companions, subject to our detectability simulations ($D(m,a)$) of the previous section. Since it remains possible that the star has a companion below our detectability limit, the likelihood of the data given a non-detection is 
\begin{equation}
L_i = 1 - \int d \ln m \, d \ln a \, D_i(m,a) \, f(m,a)
\end{equation}
Therefore, the total likelihood for a system of $N_d$ detections around $N$ stars is
\begin{align}
\mathcal{L} &=& \prod_{i=1}^{N_d} \bigg(\int d \ln m \, \int d \ln a \, f(m,a) \, p_i(m,a)\bigg) \nonumber \\
& & \times \prod_{j=1}^{N-N_d} \bigg(1 - \int d \ln m \, d \ln a \, D_j(m,a) \, f(m,a)\bigg).
\end{align}

\begin{figure}[htbp]
\centerline{\includegraphics[width=0.55\textwidth]{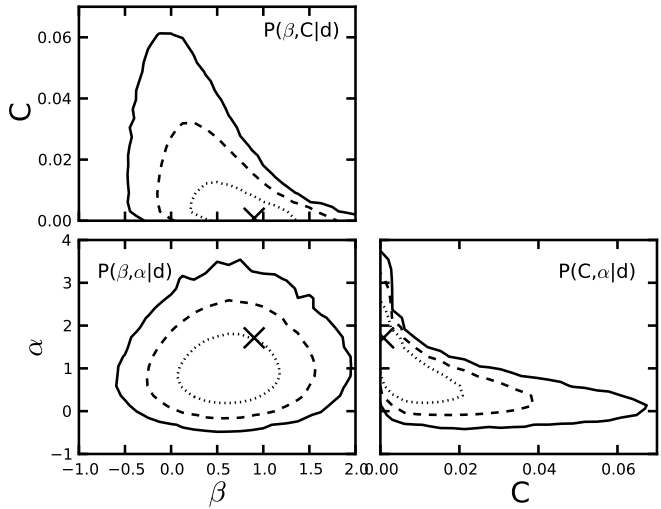}}
\caption{Covariance plots showing the likelihood of any two parameters $C$, $\alpha$, and $\beta$ given data $d$ for the full set of fifty one systems. $X$ marks on the plot correspond to the values of the parameters which correspond to the maximum likelihood value of $\mathcal{L}$.
  }
\label{alphaopik}
\end{figure}

We then vary $\alpha$, $\beta$, and our normalization factor $C$ to maximize $\mathcal{L}$. This is similar to the approach taken by \citet{cumming08}, although with their well-characterized planets, $p_i$ was approximated by these authors as a $\delta$ function. Here, we can only assume a $\delta$ function in mass and semi-major axis for the few systems with well-characterized orbits.  This approach is also functionally identical to injecting artificial planets following some distribution and matching the observed distribution to the simulated planets, in the limit as the number of injected planets approaches infinity.  We calculate constraints on $C$, $\alpha$, and $\beta$ using our likelihood function described above and the emcee package developed by \citet{foreman-mackey13}.  The distribution of acceptable values of $C$, $\alpha$, and $\beta$ are shown in Fig.~\ref{alphaopik} and listed in Table \ref{T:CAB}.  

An estimated value of the planet occurrence rate can be found by integrating $f$ over a domain of interest.  If we take a range of $1-13$~M$_{Jup}$ in mass and $1-20$ AU in orbital semi-major axes, we find a total occurrence rate of $51\pm10\%$ for our sample.  This suggests that there may be an additional $13\pm5$ companions in this range that were missed by our observations.  If we consider a smaller range of $1-10$ AU with the same mass range, the total occurrence rate is $27\% \pm6\%$, which is lower than in our previous example but also more tightly constrained.  Finally, if we consider a broader range of $0.2-13$~M$_{Jup}$ and $1-20$ AU we estimate a higher occurrence rate of $55^{+12}_{-10}\%$ with modestly increased uncertainties.  One caveat to these integrated occurrence rates is that our choice of a power law distribution may not remain accurate from Saturn mass objects all the way up to the deuterium burning limit.  We chose this formalism and domain in order to facilitate comparison to previous analyses of planet occurrence rates, which have often made the same assumptions.

We compare our ``misaligned'' sub-sample, which contains planets with misaligned and/or eccentric orbits, to our ``control'' sub-sample of well-aligned planets on circular orbits.  We find that  the companion occurrence rates in the two sub-samples are consistent within the uncertainties, suggesting that the spin-orbit alignments and eccentricities of  the short period planets are not affected by the presence of these massive companions.  All well-characterized outer planets exist in our misaligned sub-sample, so it is perhaps not surprising that the companion distribution parameters are better constrained in this sub-sample than in the control sub-sample.  

\begin{figure}[htbp]
\centerline{\includegraphics[width=0.55\textwidth]{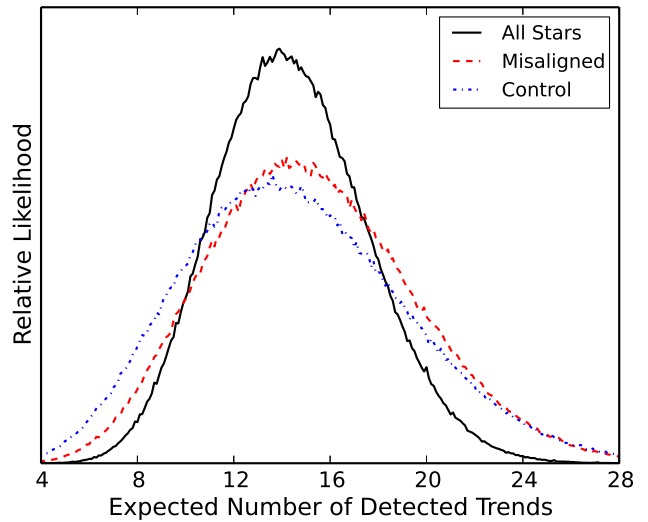}}
\caption{Expected number of detected RV accelerations, assuming our full sample of fifty-one stars is represented by the planet distribution function as calculated from the full sample (solid black), the ``misaligned'' sub-sample (red dashed; includes all systems where the inner transiting gas giant has either a non-zero eccentricity or a spin-orbit misalignment) or ``control'' sub-sample (blue dot-dashed, includes systems where the inner transiting gas giant has an apparently circular and well-aligned orbit).  The two sub-samples are consistent to within $1\sigma$, and we therefore conclude that there is no evidence to suggest that they are drawn from different populations.
  }
\label{different_populations}
\end{figure}

We investigate the consistency of the two samples by calculating the number of RV trends we would expect to observe if the giant planets in both samples were represented by the underlying planet distribution of the ``misaligned'' sub-sample.  For each value of $C$, $\alpha$, and $\beta$, we calculate the number of trends we would expect to observe, given the assumed underlying planet population and our calculated ability to detect companions around each star as a function of companion mass and semimajor axis.  We then weight each value according to the relative likelihood of that particular choice of parameters.  In the full sample, we would expect to detect $14.4 \pm 3.2$ companions based on our planet distribution function (Fig.~\ref{different_populations}).  If the full sample of stars is described by the parameterizations of the ``misaligned'' (or ``control'') sub-sample, then we would expect to detect $14.9^{+4.6}_{-4.0}$ ($14.2^{+4.9}_{-4.3}$) companions, consistent with each other and with the main sample.  We also quantify the degree of similarity between the two samples by calculating the integrated frequency for companions with masses between $1-13$~M$_{Jup}$ and orbital semi-major axes between $1-20$ AU.  We find frequencies of $46^{+12}_{-10}\%$ for the ``misaligned'' sample and $59^{+21}_{-15}\%$ for the ``control'' sample, which agree at the $1\sigma$ level.  Thus, there is no evidence to suggest these two sub-samples are drawn from different populations.  

\begin{figure}[ht]
\epsscale{2.4}
\plottwo{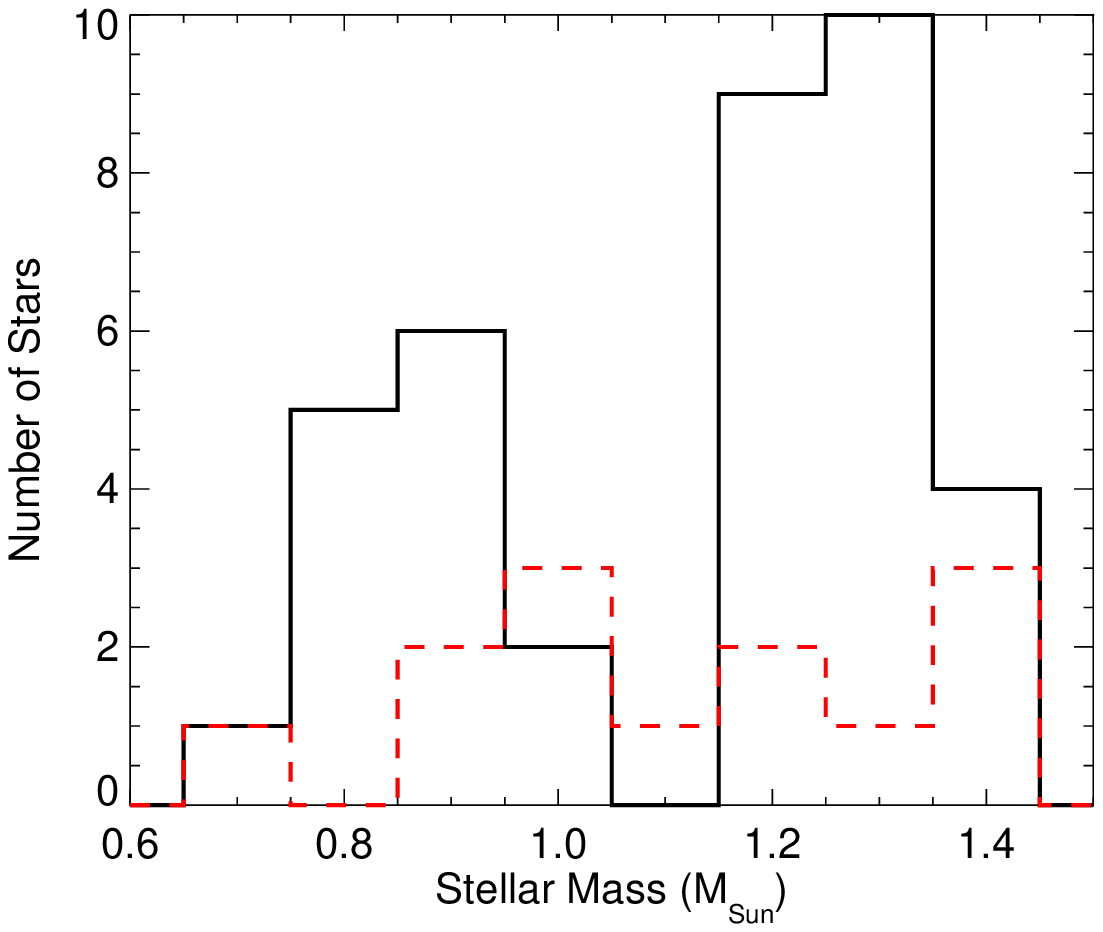}{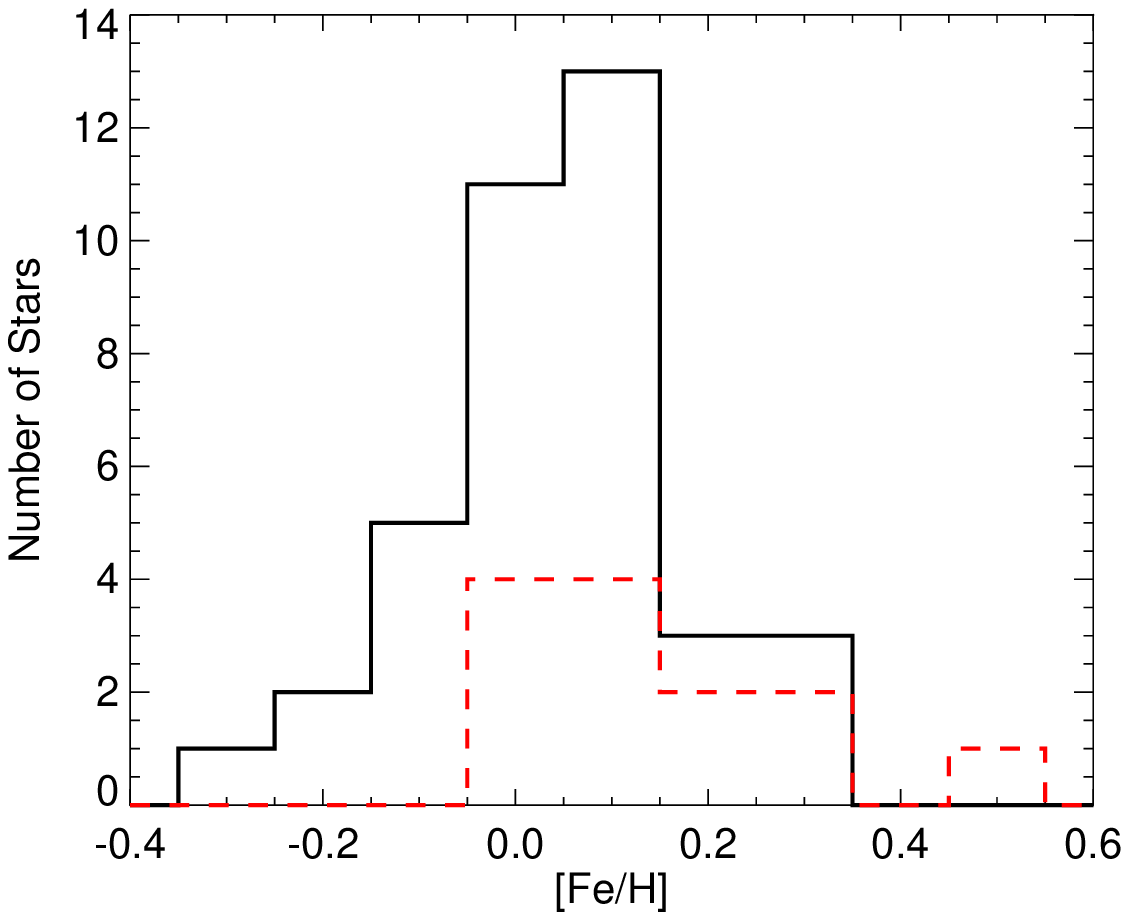}
\centering
\caption{Histogram of the stellar masses (upper panel) and metallicities (lower panel) for objects in our sample with (red dashed line) and without (black solid line) long-period radial velocity companions.  Values for individual systems and associated references are given in Table \ref{stellar_parameters}.}
\label{companion_hist}
\end{figure}

We also plotted histograms of the distribution of stellar masses and metallicities for systems with and without long-period companions (see Fig.~\ref{companion_hist}).  We tested the significance of potential correlations between companion occurrence and either stellar mass or metallicity using the Kolmogorov-Smirnov statistic, which evaluates the probability that two samples are drawn from the same parent distribution.  We find that in both cases this probability is high ($21\%$ for the two mass distributions and $67\%$ for the two metallicity distributions), indicating that there are no statistically significant correlations between companion frequency and either of these two parameters.  We note that this calculation does not take into account the variations in our sensitivity to companions for individual systems, but it is unlikely that a more complete analysis would change our conclusions on this point.

\subsection{The Multiplicity Rate of  Short Period Gas Giants As Compared to Other Planet Populations}\label{multiplicity_comparison}

We next consider how the multiplicity rate $C$ for our sample compares to those of planetary systems detected by the \textit{Kepler} transit survey and by radial velocity surveys, where we define multiplicity as the fraction of planetary systems containing more than one planet.  \citet{batalha13} report the detection of 369 systems with multiple transiting planet candidates out of a total of 1797 stars with at least one transiting planet candidate.  This corresponds to a multiplicity rate of approximately $21\%$ for the \textit{Kepler} sample.  \citet{tremaine12} perform a more detailed statistical analysis of the \textit{Kepler} planet candidate sample and find a multiplicity rate ranging between $20\%-50\%$ depending on the distribution of mutual inclinations assumed in the calculation.  Although both of these numbers are broadly consistent with our multiplicity rate, we note that the characteristics of the \textit{Kepler} candidate multi-planet systems are dramatically different than in our sample.  Our systems consist of a short-period, gas giant planet with another massive companion on a very long period (typically several years or more) orbit.  In contrast, the \textit{Kepler} candidate multi-planet systems typically consist of tightly-packed sets of low-mass (smaller than Neptune) planets on orbits less than 100 days (e.g., Latham et al. 2011, Fabrycky et al. 2012, Steffen et al. 2013).  \citet{steffen12} used the \textit{Kepler} data set to demonstrate that hot Jupiter candidates are notably lacking in nearby, low-mass companions.  For the majority of the \textit{Kepler} systems, the frequency of massive long-period companions is unknown; these planets are unlikely to transit and radial velocity follow-up is challenging for most \textit{Kepler} targets.

The sample of radial velocity planets provides a better basis for comparison, as it includes many systems with long-term radial velocity monitoring capable of detecting massive companions on long-period orbits \citep[e.g.,][]{fischer01,wright07,wright09}.  \citet{tremaine12} find a total of 162 single-planet systems and 33 multi-planet systems detected orbiting FGK dwarf stars as of August 2010.  This corresponds to a multiplicity fraction of $17\%\pm3\%$, but this number only includes planets with fully resolved orbits.  \citet{wright09} carry out a similar analysis including radial velocity accelerations, where they exclude accelerations in systems where there is a known stellar companion.  They find that $14\%$ of the 205 planetary systems known at the time have multiple confirmed planets, with another $14\%$ showing evidence for an additional distant companion.  They also note that this number is most likely an underestimate, as the occurrence rates for planets increase sharply towards smaller masses and radii \citep[e.g.,][]{howard10,petigura13} where their survey has a high level of incompleteness.  Our survey has a similar lack of sensitivity to very small, distant planets, and we find evidence for at least one companion around $27\%$ of our target stars.  Even after accounting for planets we might have missed, our multiplicity rate of $51\pm10\%$ for large companions to the transiting planets in our sample is still in reasonably good agreement with their result.  \citet{wright09} also note that the observed excess of giant planets at short orbital periods (the ``three-day pile-up") disappears for multi-planet systems, indicating that nearby gas giant companions are rare in systems with hot Jupiters.  Our results support this finding, as there is no evidence for any gas giant companions interior to 1 AU for the systems in our survey.

\section{Conclusions}

We combine K band AO imaging and long-term radial velocity monitoring to place constraints on the presence of distant, massive companions to a sample of 51 transiting gas giant planets.  We find evidence for fifteen companions in fourteen systems, including new detections in six systems:  HAT-P-10, HAT-P-22, HAT-P-29, HAT-P-32, WASP-10, and XO-2.  For the HAT-P-10 system, we conclude that the observed radial velocity trend is likely due to a directly imaged stellar companion at larger separations and therefore exclude it from our subsequent analysis of planetary mass companions.  We also detect a trend in the HAT-P-11 system that is well-correlated with the \caii~emission line strength, suggesting that it is likely the result of stellar activity and not a long-period companion.  We find no evidence for companions in three systems with previously reported radial velocity trends, including: GJ 436, HAT-P-31, and HAT-P-34.

For the systems with radial velocity accelerations consistent with the presence of a long-period companion, we place upper limits on the mass and period of the companion using K band AO images.  One companion (HAT-P-13c) has a fully resolved orbit, while three additional companions (HAT-P-17c, WASP-8c, and WASP-34c) display noticeable curvature in their radial velocity accelerations.  For the remaining systems, the linearity of the trend places a lower limit on the mass and semi-major axis of the companion's orbit.  Combining these two constraints, we find that the companions in these systems typically have masses constrained to lie between $1-500$ M$_{Jup}$ and orbital semi-major axes between $1-75$ AU.  A significant majority of these companions are constrained to have minimum masses that are larger than those of the transiting planets in these systems.  Although we cannot evaluate the plausibility of specific dynamical scenarios without more precise knowledge of the masses and orbital configurations of these outer companions, we note that a recent study by \citet{teyssandier13} found that inward migration and spin-orbit misalignment of a Jupiter-mass planet initially located at 5 AU was most likely when the outer companion had a mass at least twice that of the Jupiter and was located in an eccentric, misaligned orbit with a semi-major axis between $50-150$ AU.  Many of our companions have semi-major axis ranges that extend beyond 10 AU, and we expect that additional radial velocity monitoring will continue to improve our constraints on their orbital periods.

We estimate a total occurrence rate of $51\pm10\%$ for companions with masses between $1-13$~M$_{Jup}$ and orbital semi-major axes between $1-20$ AU in our sample.  We find no statistically significant difference between the frequency of companions in systems with misaligned or eccentric orbits and those with well-aligned, circular orbits.  This is still consistent with the hypothesis that spin-orbit misalignments are the result of dynamical interactions with a distant outer companion, as \citet{albrecht12b} have proposed that all hot Jupiters are initially misaligned and that stellar tides bring a subset of the sample back into alignment.  However, the exact nature of this tidal realignment is still debated \citep{lai12,rogers13,valsecchi14}.  If tides are the cause of the aligned systems we would expect companions to be common in all hot Jupiter systems, regardless of their present-day orbital alignment.  We also find no evidence for any statistically significant correlations between companion occurrence and either the mass or metallicity of the host star. 

The total companion frequency in our sample is comparable to the multiplicity rates from the \textit{Kepler} mission and from radial velocity surveys.  However, the compact, low-mass candidate multi-planet systems detected by \textit{Kepler} can have up to five sub-Neptune-sized planets with orbital periods less than fifty days.  In contrast to these systems, the companions we find in our sample all have periods of a year or longer and masses that are invariably larger than those of the inner transiting  gas giants.  HAT-P-13 is particularly noteworthy in our sample as the only system with two massive outer companions, resulting in potentially interesting dynamical interactions \citep[e.g.,][]{batygin09,becker13}.  

We note that the presence of distant massive stellar companions, such as those around HAT-P-7 \citep{narita12}, HAT-P-10 (Ngo et al. in prep), HAT-P-32 \citep{adams13}, and WASP-8 \citep{queloz10}, may also play an important role in driving the dynamical evolution of these systems.  In the case of HAT-P-7, HAT-P-32, and WASP-8, the directly imaged stellar companion is too distant to explain the measured radial velocity trend, indicating that these hot Jupiters have not one but two massive outer companions.  In our next paper we will present the results of a comprehensive AO survey of all 51 systems in our sample, which will allow us to evaluate the frequency of stellar companions on wide-separation orbits beyond approximately 50 AU  This comprehensive study will complete our picture of these systems and provide an invaluable test of competing theories for the underlying cause of hot Jupiter migration and spin orbit misalignments.

\acknowledgements
We thank Jason Wright and Joshua Winn for their thoughtful input on early drafts of this paper.  This work was based on observations at the W. M. Keck Observatory granted by the University of Hawaii, the University of California, and the California Institute of Technology. We thank the observers who contributed to the measurements reported here and acknowledge the efforts of the Keck Observatory staff. We extend special thanks to those of Hawaiian ancestry on whose sacred mountain of Mauna Kea we are privileged to be guests.  B.T.M. is supported by the National Science Foundation Graduate Research Fellowship under Grant No. DGE$-$1144469.  J.A.J. and G.B. are supported by grants from the David and Lucille Packard Foundation and the Alfred P. Sloan Foundation.  P.S.M. acknowledges support for this work from the Hubble Fellowship Program, provided by NASA through Hubble Fellowship grant HST-HF-51326.01-A awarded by the STScI, which is operated by the AURA, Inc., for NASA, under contract NAS 5-26555.  S.H. is supported by an NSF Astronomy and Astrophysics Postdoctoral Fellowship under award AST-1203023.  G.B. acknowledges support from grant NSFAST-1108686.

{\it Facilities:} \facility{Keck:I (HIRES)}, \facility{Keck:II (NIRC2)}


\begin{thebibliography}{}

\bibitem[Adams et al.(2013)]{adams13} Adams, E. R., Dupree, A. K., Kulesa, C., \& McCarthy, D. 2013, \apj, 146, 9

\bibitem[Albrecht et al.(2011)]{albrecht11} Albrect, S. et al. 2011, \apj, 738, 50 
\bibitem[Albrecht et al.(2012a)]{albrecht12a} Albrecht, S. et al. 2012a, \apj, 744, 189 
\bibitem[Albrecht et al.(2012b)]{albrecht12b} Albrecht, S. et al. 2012b, \apj, 757, 18 
\bibitem[Albrecht et al.(2013)]{albrecht13} Albrecht, S. et al. 2013, \apj, 771, 11
\bibitem[Anderson et al.(2010)]{anderson10} Anderson, D. R. et al. 2010, \apj, 709, 159 
\bibitem[Anderson et al.(2011)]{anderson11} Anderson, D. R. et al. 2011, \aap, 534, A16 
\bibitem[Anderson et al.(2013)]{anderson13} Anderson, D. R. et al. 2013, \mnras, 430, 3422
\bibitem[Bakos et al.(2009a)]{bakos09a} Bakos, G. \' A. et al. 2009a, \apj, 696, 1950 
\bibitem[Bakos et al.(2009b)]{bakos09b} Bakos, G. \' A. et al. 2009b, \apj, 707, 446 
\bibitem[Bakos et al.(2010)]{bakos10} Bakos, G. \' A. et al. 2010, \apj, 710, 1724 
\bibitem[Bakos et al.(2011)]{bakos11} Bakos, G. \' A. et al. 2011, \apj, 742, 116 
\bibitem[Bakos et al.(2012)]{bakos12} Bakos, G. \' A. et al. 2012, \apj, 144, 19 
\bibitem[Baraffe et al.(1998)]{baraffe98} Baraffe, I., Chabrier, G., Allard, F., \& Hauschildt, P. H. 1998, \aap, 337, 403
\bibitem[Barclay et al.(2012)]{barclay12} Barclay, T. et al. 2012, \apj, 761, 53 
\bibitem[Barnes et al.(2013)]{barnes13} Barnes, J., van Eyken, J. C., Jackson, B. K., Ciardi, D. R., \& Fortney, J. J. 2013, \apj, 774, 53
\bibitem[Barros et al.(2011)]{barros11} Barros, S. C. C. et al. 2011, \aap, 525, A54 
\bibitem[Barros et al.(2013)]{barros13} Barros, S. C. C. et al. 2013, \mnras, 430, 3032 
\bibitem[Batalha et al.(2013)]{batalha13} Batalha, N. et al. 2013, \apjs, 204, 24
\bibitem[Batygin et al.(2009)]{batygin09} Batygin, K., Bodenheimer, P., \& Laughlin, G. 2009, \apj, 704, L49
\bibitem[Batygin(2012)]{batygin12} Batygin, K. 2012, Nature, 491, 418
\bibitem[Batygin \& Adams(2013)]{batygin13} Batygin, K. \& Adams, F. C. 2013, \apj in press, arXiv:1310.2179
\bibitem[Bechter et al.(2013)]{bechter13} Bechter, E. B. et al. 2013, \apj submitted, arXiv:1307.6857
\bibitem[Becker \& Batygin(2013)]{becker13} Becker, J. C. \& Batygin, K. 2013, \apj, 778, 100
\bibitem[Beerer et al.(2011)]{beerer11} Beerer, I. M. et al. 2011, \apj, 727, 23
\bibitem[Blecic et al.(2012)]{blecic12} Blecic, J. et al. 2012, \apj submitted, arXiv:1111.2363
\bibitem[Bonfils et al.(2005)]{bonfils05} Bonfils, X. et al. 2005, \aap, 442, 635
\bibitem[Bowler et al.(2010)]{bowler10} Bowler, B. P. et al. 2010, \apj, 709, 396
\bibitem[Brown et al.(2012a)]{brown12} Brown, D. J. A. et al. 2012a, \mnras, 423, 1503 
\bibitem[Brown et al.(2012b)]{brown12b} Brown, D. J. A. et al. 2012b, \apj, 760, 139 
\bibitem[Buchhave et al.(2010)]{buchhave10} Buchhave, L. A., et al. 2010, \apj, 720, 1118 
\bibitem[Buchhave et al.(2011)]{buchhave11} Buchhave, L. A., et al. 2011, \apj, 733, 116 
\bibitem[Burke et al.(2007)]{burke07} Burke, C. J. et al. 2007, \apj, 671, 2115 
\bibitem[Burke et al.(2008)]{burke08} Burke, C. J. et al. 2008, \apj, 686, 1331 
\bibitem[Cameron et al.(2007)]{cameron07} Cameron, A. C. et al. 2007, \mnras, 375, 951
\bibitem[Campo et al.(2011)]{campo11} Campo, C. J. et al. 2011, \apj, 727, 125
\bibitem[Carter et al.(2009)]{carter09} Carter, J. A., Winn, J. N., Gilliland, R., \& Holman, M. J. 2009, \apj, 696, 241 
\bibitem[Chatterjee et al.(2008)]{chatterjee08} Chatterjee, S., Ford, E. B., Matsumura, S., \& Rasio, F. A. 2008, \apj, 686, 580 
\bibitem[Christian et al.(2009)]{christian09} Christian, D. J. et al. 2009, \mnras, 392, 1585 
\bibitem[Christiansen et al.(2010)]{christiansen10} Christiansen, J. L. et al. 2010, \apj, 710, 97
\bibitem[Crepp et al.(2012)]{crepp12} Crepp, J. R. et al. 2012, \apj, 761, 39
\bibitem[Croll et al.(2010a)]{croll10a} Croll, B. et al. 2010a, \apj, 717, 1084
\bibitem[Croll et al.(2010b)]{croll10b} Croll, B. et al. 2010b, \apj, 718, 920
\bibitem[Cubillos et al.(2013)]{cubillos13} Cubillos, P. et al. 2013, \apj, 768, 42
\bibitem[Cumming et al.(2008)]{cumming08} Cumming, A. et al. 2008, \pasp, 120, 531
\bibitem[Dawson et al.(2012)]{dawson12} Dawson, R. I., Murray-Clay, R. A., \& Johnson, J. A. 2012, \apj submitted, arXiv:1211:0554
\bibitem[Doyle et al.(2013)]{doyle13} Doyle, A. P. et al. 2013, \mnras, 428, 3164
\bibitem[Duquennoy \& Mayor(1991)]{duquennoy91} Duquennoy, A. \& Mayor, M. 1991, \aap, 248, 485
\bibitem[Eastman et al.(2013)]{eastman13} Eastman, J., Gaudi, B. S., \& Agol, E. 2013, PASP, 125, 83
\bibitem[Fabrycky \& Tremaine(2007)]{fabrycky07} Fabrycky, D. \& Tremaine, S. 2007, \apj, 669, 1298
\bibitem[Fischer et al.(2001)]{fischer01} Fischer, D. A. et al. 2001, \apj, 551, 1107
\bibitem[Ford(2006)]{ford06} Ford, E. B. 2006, \apj, 642, 505
\bibitem[Foreman-Mackey(2013)]{foreman-mackey13} Foreman-Mackey, D., Hogg, D. W., Lang, D., \& Goodman, J. 2013, \pasp, 125, 306
\bibitem[Fressin et al.(2010)]{fressin10} Fressin, F. et al. 2010, \apj, 711, 374
\bibitem[Fulton et al.(2013)]{fulton13} Fulton, B. J. et al. 2013, \apj, 772, 80
\bibitem[Gelman et al.(2003)]{gelman03} Gelman, A., Carlin, J. B., Stern, H. S., \& Rubin, D. B. 2003, Bayesian Data Analysis, 2nd edn.(Chapban and Hall) 
\bibitem[Gibson et al.(2008)]{gibson08} Gibson, N. P. et al. 2008, \aap, 492, 603 
\bibitem[Girardi et al.(2002)]{girardi02} Girardi, L. et al. 2002, \aap, 391, 195
\bibitem[Goldreich \& Tremaine(1980)]{goldreich80} Goldreich, P., \& Tremaine, S. 1980, \apj, 241, 425
\bibitem[Goldreich \& Sari(2003)]{goldreich03} Goldreich, P. \& Sari, R. 2003, \apj, 585, 1024
\bibitem[Hartman et al.(2009)]{hartman09} Hartman, J. D. et al. 2009, \apj, 706, 785 
\bibitem[Hartman et al.(2011a)]{hartman11a} Hartman, J. D. et al. 2011a, \apj, 726, 52 
\bibitem[Hartman et al.(2011b)]{hartman11b} Hartman, J. D. et al. 2011b, \apj, 728, 138 
\bibitem[Hartman et al.(2011c)]{hartman11c} Hartman, J. D. et al. 2011c, \apj, 742, 59 
\bibitem[Hebb et al. (2009a)]{hebb09a} Hebb, L. et al. 2009a, \apj, 693, 1920
\bibitem[Hebb et al. (2009b)]{hebb09b} Hebb, L. et al. 2009b, \apj, 708, 224
\bibitem[Hebrard et al.(2008)]{hebrard08} Hebrard, G., et al. 2008, \aap, 488, 763
\bibitem[Hebrard et al.(2011)]{hebrard11} Hebrard, G., et al. 2011, \aap, 527, L11
\bibitem[Hellier et al.(2008)]{hellier08} Hellier, C. et al. 2008, \apj, 690, L89 
\bibitem[Hellier et al.(2009)]{hellier09} Hellier, C. et al. 2009, Nature, 460, 1098 
\bibitem[Hellier et al.(2011)]{hellier11} Hellier, C. et al. 2011, \apj, 730, L31 
\bibitem[Hirano et al.(2011)]{hirano11} Hirano, T. et al. 2011, PASJ, 63, L57 
\bibitem[Hirano et al.(2012)]{hirano12} Hirano, T. et al. 2012, \apj, 759, L36
\bibitem[Holman et al.(1997)]{holman97} Holman, M., Touma, J., \& Tremaine, S. 1997, Nature, 386, 254
\bibitem[Holman et al.(2006)]{holman06} Holman, M. J. et al. 2006, \apj, 652, 1715
\bibitem[Howard et al.(2009)]{howard09} Howard, A. W. et al. 2009, \apj, 696, 75
\bibitem[Howard et al.(2010)]{howard10} Howard, A. W. et al. 2010, Science, 330, 653
\bibitem[Howard et al.(2012)]{howard12} Howard, A. W. et al. 2012, \apj, 749, 134 
\bibitem[Huber et al.(2013)]{huber13} Huber, D. et al. 2013, Science, 342, 331
\bibitem[Husnoo et al.(2011)]{husnoo11} Husnoo, N. et al. 2011, \mnras, 413, 2500
\bibitem[Husnoo et al.(2012)]{husnoo12} Husnoo, N. et al. 2012, \mnras, 422, 3151
\bibitem[Husser et al.(2013)]{husser13} Husser, T.-O. et al. 2013, \aap, 553, A6
\bibitem[Isaacson \& Fischer(2010)]{isaacson10} Isaacson, H. \& Fischer, D. 2010, \apj, 725, 875
\bibitem[Johns-Krull et al.(2008)]{johnskrull08} Johns-Krull, C. M. et al. 2008, \apj, 677, 657
\bibitem[Johnson et al.(2009a)]{johnson09} Johnson, J. A. et al. 2009a, \pasp, 121, 1104 
\bibitem[Johnson et al.(2009b)]{johnson09b} Johnson, J. A. et al. 2009b, \apj, 692, L100 
\bibitem[Johnson et al.(2010)]{johnson10} Johnson, J. A. et al. 2010, \pasp, 122, 149
\bibitem[Johnson et al.(2011)]{johnson11} Johnson, J. A. et al. 2011, \apj, 735, 24 
\bibitem[Joshi et al.(2009)]{joshi09} Joshi, Y. C. et al. 2009, \mnras, 392, 1532 
\bibitem[Juric \& Tremaine(2008)]{juric08} Juric, M. \& Tremaine, S. 2008, \apj, 686, 603
\bibitem[Kaib et al.(2011)]{kaib11} Kaib, N. A., Raymond, S. N., \& Duncan, M. J. 2011, \apj, 742, L24
\bibitem[Kane \& Raymond(2014)]{kane14a} Kane, S. R. \& Raymond, S. N. 2014, \apj submitted, arXiv:1401.7998
\bibitem[Kane et al.(2014)]{kane14b} Kane, S. R. et al. 2014, \apj submitted, arXiv:1401.1544
\bibitem[Kipping et al.(2010)]{kipping10} Kipping, D. M. et al. 2010, \apj, 725, 2017 
\bibitem[Kipping et al.(2011)]{kipping11} Kipping, D. M. et al. 2011, \apj, 142, 95 
\bibitem[Kipping(2013)]{kipping13} Kipping, D. M. 2013, \mnras, 434, L51
\bibitem[Knutson et al.(2009)]{knutson09} Knutson, H. A. et al. 2009, \apj, 691, 866
\bibitem[Knutson et al.(2010)]{knutson10} Knutson, H. A. et al. 2010, \apj, 720, 1569
\bibitem[Knutson et al.(2011)]{knutson11} Knutson, H. A. et al. 2011, \apj, 735, 27
\bibitem[Kov\' acs et al.(2007)]{kovacs07} Kov\' acs, G. et al. 2007, \apj, 670, L41
\bibitem[Kov\' acs et al.(2010)]{kovacs10} Kov\' acs, G. et al. 2010, \apj,724, 866 
\bibitem[Lewis et al.(2013)]{lewis13} Lewis, N. K. et al. 2013, \apj, 766, 95
\bibitem[Lai(2012)]{lai12} Lai, D. 2012, \mnras 423, 486
\bibitem[Li et al.(2013)]{li13} Li, G., Naoz, S., Kocsis, B., \& Loeb, A. 2013, Nature submitted, arXiv:1310.6044
\bibitem[Lin \& Papaloizou(1986)]{lin86} Lin, D. N. C. \& Papaloizou, J. C. B., 1986, \apj, 309, 846
\bibitem[Lin et al.(1996)]{lin96} Lin, D. N. C., Bodenheimer, P., \& Richardson, D. C. 1996, Nature, 380, 606
\bibitem[Lister et al.(2009)]{lister09} Lister, T. A. et al. 2009, \apj, 703, 752 
\bibitem[Lithwick \& Wu(2013)]{lithwick13} Lithwick, Y. \& Wu, Y. 2013, PNAS submitted, arXiv:1311.1214
\bibitem[Liu et al.(2002)]{liu02} Liu, M. C. et al. 2002, \apj, 571, 519
\bibitem[Machalek et al.(2009)]{machalek09} Machalek, P. et al. 2009, \apj, 701, 514
\bibitem[Machalek et al.(2010)]{machalek10} Machalek, P. et al. 2010, \apj, 711, 111
\bibitem[Maciejewski et al.(2011a)]{maciejewski11a} Maciejewski, G. et al. 2011a, \aap, 528, A65 
\bibitem[Maciejewski et al.(2011b)]{maciejewski11b} Maciejewski, G. et al. 2011b, Acta Astronomica, 61, 1
\bibitem[Madhusudhan \& Winn(2009)]{madhu09} Madhusudhan, N. \& Winn, J. N. 2009, \apj, 693, 784
\bibitem[Malmberg et al.(2007)]{malmberg07} Malmberg, D., Davies, M. B., \& Chambers, J. E. 2007, \mnras, 377, L1
\bibitem[Mancini et al.(2013a)]{mancini13} Mancini, L. et al. 2013a, \aap, 551, A11
\bibitem[Mancini et al.(2013b)]{mancini13b} Mancini, L. et al. 2013b, \mnras, 436, 2 
\bibitem[Mandushev et al.(2007)]{mandushev07} Mandushev, G. et al. 2007, \apj, 667, L195 
\bibitem[Maness et al.(2007)]{maness07} Maness, H. L. et al. 2007, \pasp, 119, 90
\bibitem[Marcy \& Butler(1992)]{marcy92} Marcy, G. W. \& Butler, R. P. 1992, \pasp, 104, 270
\bibitem[Maxted et al.(2010)]{maxted10} Maxted, P. F. L. et al. 2010, \apj, 140, 2007 
\bibitem[Maxted et al.(2013)]{maxted13} Maxted, P. F. L. et al. 2013, \apj, 428, 2645 
\bibitem[McCullough et al.(2008)]{mccullough08} McCullough, P. R. et al. 2008, \apj submitted, arXiv:0805.2921
\bibitem[Montet et al.(2013)]{montet13} Montet, B. T. et al. 2013, \apj, 781, 28
\bibitem[Morris et al.(2013)]{morris13} Morris, B. M., Mandell, A. M., \& Deming, D. 2013, \apj, 764, L22
\bibitem[Morton \& Johnson(2011)]{morton11} Morton, T. D., \& Johnson, J. A. 2011, \apj, 729, 138
\bibitem[Moutou et al.(2011)]{moutou11} Moutou, C. et al. 2011, \aap, 533, A113 
\bibitem[Nagasawa et al.(2008)]{nagasawa08} Nagasawa, M., Ida, S., \& Bessho, T. 2008, ApJ, 678, 498
\bibitem[Naoz et al.(2012)]{naoz12} Naoz, S. et al. 2012, \apj, 754, L36
\bibitem[Narita et al.(2010a)]{narita10a} Narita, N. et al. 2010a, PASJ, 62, 653 
\bibitem[Narita et al.(2010b)]{narita10} Narita, N. et al. 2010b, PASJ, 62, L61 
\bibitem[Narita et al.(2011)]{narita11} Narita, N. et al. 2011, PASJ, 63, L67 
\bibitem[Narita et al.(2012)]{narita12} Narita, N. et al. 2012, PASJ, 64, L7 
\bibitem[Nikolov et al.(2012)]{nikolov12} Nikolov, N. et al. 2012, \aap, 539, A159
\bibitem[Noyes et al.(2008)]{noyes08} Noyes, R. W. et al. 2008, \apj, 673, L79
\bibitem[Nymeyer et al.(2010)]{nymeyer10} Nymeyer, S. et al. 2010, \apj, 742, 35
\bibitem[O'Donovan et al.(2007)]{odonovan07} O'Donovan, F. T. et al. 2007, \apj, 663, L37
\bibitem[O'Donovan et al.(2010)]{odonovan10} O'Donovan, F. T. et al. 2010, \apj, 710, 1551
\bibitem[P\' al et al.(2008)]{pal08} P\' al, A. et al. 2008, \apj, 680, 1450
\bibitem[P\' al et al.(2009)]{pal09} P\' al, A. et al. 2009, \apj, 700, 783 
\bibitem[P\' al et al.(2010)]{pal10} P\' al, A. et al. 2010, \mnras, 401, 2665 
\bibitem[Petigura et al.(2013)]{petigura13} Petigura, E. A., Howard, A. W., \& Marcy, G. W. 2013, PNAS, 110, 48
\bibitem[Pollacco et al.(2008)]{pollacco08} Pollacco, D. et al. 2008, \mnras, 385, 1576
\bibitem[Pont et al.(2011)]{pont11} Pont, F., Husnoo, N., Mazeh, T., \& Fabrycky, D. 2011, \mnras, 414, 1278
\bibitem[Queloz et al.(2010)]{queloz10} Queloz, D. et al. 2010, \aap, 517, L1
\bibitem[Quinn et al.(2013)]{quinn13} Quinn, S. N. et al. 2013, \apj submitted, arXiv:1310.7328
\bibitem[Raghavan et al.(2010)]{raghavan10} Raghavan, D. et al. 2010, \apjs, 190, 1
\bibitem[Rogers \& Lin(2013)]{rogers13} Rogers, T. M. \& Lin, D. N. C. 2013, \apj 769, L10
\bibitem[Sada et al.(2012)]{sada12} Sada, P. V. et al. 2012, \pasp, 124, 212
\bibitem[Sanchis-Ojeda et al.(2011)]{sanchis11} Sanchis-Ojeda, R. et al. 2011, \apj, 733, 127 
\bibitem[Sanchis-Ojeda et al.(2012)]{sanchis12} Sanchis-Ojeda, R. et al. 2012, Nature, 487, 449
\bibitem[Simpson et al.(2010)]{simpson10} Simpson, E. K. et al. 2010, \mnras, 405, 1867
\bibitem[Simpson et al.(2011)]{simpson11} Simpson, E. K. et al. 2011, \mnras, 414, 3023 
\bibitem[Sing et al.(2012)]{sing12} Sing, D. K. et al. 2012, \mnras, 426, 1663 
\bibitem[Smalley et al.(2011)]{smalley11} Smalley, B. et al. 2011, \aap, 526, A130 
\bibitem[Smith et al.(2012)]{smith12} Smith, A. M. S. et al. 2012, \aap, 545, A93 
\bibitem[Southworth et al.(2012a)]{southworth12} Southworth, J., Bruni, I., Mancini, L., \& Gregorio, J. 2012a, \mnras, 420, 2580 
\bibitem[Southworth et al.(2012b)]{southworth12b} Southworth, J. et al. 2012b, \mnras, 426, 1291 
\bibitem[Southworth et al.(2012c)]{southworth12c} Southworth, J. et al. 2012c, \mnras, 426, 1338 
\bibitem[Southworth et al.(2013)]{southworth13} Southworth, J. et al. 2013, \mnras, 434, 1300 
\bibitem[Sozzetti et al.(2009)]{sozzetti09} Sozzetti, A. et al. 2009, \apj, 691, 1145 
\bibitem[Steffen et al.(2012)]{steffen12} Steffen, J. H. et al. 2012, PNAS, 109, 7982
\bibitem[Steffen et al.(2013)]{steffen13} Steffen, J. H. et al. 2013, \mnras, 428, 1077
\bibitem[Stevenson et al.(2010)]{stevenson10} Stevenson, K. B. et al. 2010, Nature, 464, 1161
\bibitem[Stevenson et al.(2012)]{stevenson12} Stevenson, K. B. et al. 2012, \apj, 754, 136
\bibitem[Street et al.(2010)]{street10} Street, R. A. et al. 2010, \apj, 720, 337 
\bibitem[Tanaka et al.(2002)]{tanaka02} Tanaka, H., Takeuchi, T., \& Ward, W. R. 2002, \apj, 565, 1257
\bibitem[Ter Braak(2006)]{braak06} Ter Braak, C. 2006, Statistics and Computing 
\bibitem[Teyssandier et al.(2013a)]{teyssandier13a} Teyssandier, J., Terquem, C., \& Papaloizou, J. C. B. 2013, \mnras 428, 658
\bibitem[Teyssandier et al.(2013b)]{teyssandier13} Teyssandier, J., Naoz, S., Lizarraga, I., \& Rasio, F. A. 2013, \apj 779, 166
\bibitem[Todorov et al.(2011)]{todorov11} Todorov, K. O. et al. 2011, \apj, 746, 111
\bibitem[Todorov et al.(2013)]{todorov13} Todorov, K. O. et al. 2013, \apj, 770, 102
\bibitem[Torres(1999)]{torres99} Torres, G. et al. 1999, \pasp, 111, 169
\bibitem[Torres et al.(2008)]{torres08} Torres, G. et al. 2008, \apj, 677, 1324
\bibitem[Torres et al.(2010)]{torres10} Torres, G. et al. 2010, \apj, 715, 458
\bibitem[Torres et al.(2012)]{torres12} Torres, G. et al. 2012, \apj, 757, 161
\bibitem[Tregloan-Reed et al.(2012)]{tregloan12} Tregloan-Reed, J., Southworth, J., \& Tappert, C. 2012, \mnras, 428, 3671 
\bibitem[Tremaine \& Dong(2012)]{tremaine12} Tremaine, S. \& Dong, S. 2012, \apj, 143, 94
\bibitem[Triaud et al.(2010)]{triaud10} Triaud, A. H. M. J., et al. 2010, \aap, 524, A25
\bibitem[Tripathi et al.(2010)]{tripathi10} Tripathi, A. et al. 2010, \apj, 715, 421 
\bibitem[Tsang et al.(2013)]{tsang13} Tsang, D., Turner, N. J., \& Cumming, A. 2013, \apj submitted, arXiv:1310.8627
\bibitem[Turner et al.(2013)]{turner13} Turner, J. D. et al. 2013, \mnras, 428, 678
\bibitem[Valenti et al.(1995)]{valenti95} Valenti, J. A., Butler, R. P., \& Marcy, G. W. 1995, \pasp, 107, 966
\bibitem[Valsecchi \& Rasio(2014)]{valsecchi14} Valsecchi, F. \& Rasio, F. A. 2014, \apj submitted, arXiv:1402.3857
\bibitem[Van Eyken et al.(2012)]{vaneyken12} Van Eyken, J. C. et al. 2012, \apj, 755, 42
\bibitem[Van Eylen et al.(2012)]{vaneylen12} Van Eylen, V. et al. 2012, Astron. Nachr. 333, 1088
\bibitem[Vogt et al.(1994)]{vogt94} Vogt, S. S., et al. 1994, Proc. SPIE Instr. in Astronomy VIII, 2198, 362
\bibitem[Von Braun et al.(2012)]{vonbraun12} von Braun, K. et al. 2012, \apj, 753, 171
\bibitem[Ward(1997)]{ward97} Ward, W. R. 1997, Icarus, 126, 261
\bibitem[West et al.(2009a)]{west09} West, R. G. et al. 2009a, \aj, 137, 4834 
\bibitem[West et al.(2009b)]{west09b} West, R. G. et al. 2009b, \aap, 502, 395 
\bibitem[Wheatley et al.(2010)]{wheatley10} Wheatley, P. J. et al. 2010, \apj submitted, arXiv:1004.0836
\bibitem[Wilson et al.(2008)]{wilson08} Wilson, D. M. et al. 2008, \apj, 675, L113 
\bibitem[Winn et al.(2005)]{winn05} Winn, J. N. et al. 2005, \apj, 631, 1215
\bibitem[Winn et al.(2008a)]{winn08} Winn, J. N. et al. 2008a, \apj, 682, 1283 
\bibitem[Winn et al.(2008b)]{winn08b} Winn, J. N. et al. 2008b, \apj, 683, 1076 
\bibitem[Winn et al.(2009a)]{winn09a} Winn, J. N. et al. 2009a, \apj, 700, 302 
\bibitem[Winn et al.(2009b)]{winn09} Winn, J. N. et al. 2009b, \apj, 703, L99 
\bibitem[Winn et al.(2010a)]{winn10a} Winn, J. N., et al. 2010a, \apj, 718, 575 
\bibitem[Winn et al.(2010b)]{winn10b} Winn, J. N., Fabrycky, D., Albrecht, S., \& Johnson, J. A. 2010b, \apj, 718, L145
\bibitem[Winn et al.(2010c)]{winn10c} Winn, J. N., et al. 2010c, \apj, 723, L223
\bibitem[Winn et al.(2011)]{winn11} Winn, J. N., et al. 2011, \apj, 141, 63 
\bibitem[Wizinowich(2000)]{wizinowich00} Wizinowich, P. et al. 2000, \pasp, 112, 769 
\bibitem[Wright et al.(2004)]{wright04} Wright, J. T., Marcy, G. W., Butler, R. P., \& Vogt, S S. 2004, \apjs, 152, 261
\bibitem[Wright et al.(2007)]{wright07} Wright, J. T. et al. 2007, \apj, 657, 533
\bibitem[Wright et al.(2009)]{wright09} Wright, J. T. et al. 2009, \apj, 693, 1084
\bibitem[Wu \& Murray(2003)]{wu03} Wu, Y. \& Murray, M. 2003, \apj, 589, 605
\bibitem[Wu \& Lithwick(2011)]{wu11} Wu, Y., \& Lithwick, Y. 2010, \apj, 735, 109

\end{thebibliography}
\end{document}